\documentclass[twocolumn]{aastex62}

\newcommand{\gadget}{\textsc{Gadget-3}}
\newcommand{\sphgal}{\textsc{SPHGal}}
\newcommand{\subfind}{\textsc{Subfind}}

\newcommand{\griffin}{\textsc{griffin}}

\graphicspath{{./}{figures/}}

\received{10 Aug 2020}
\accepted{8 Oct 2020}
\submitjournal{ApJ}

\shorttitle{}
\shortauthors{Lah\'{e}n et al.}

\begin{document}

\title{Structure and rotation of young massive star clusters in a simulated dwarf starburst}

\correspondingauthor{Natalia Lah\'{e}n}
\email{natalia.lahen@helsinki.fi}

\author[0000-0003-2166-1935]{Natalia Lah\'{e}n}
\affil{Department of Physics, University of Helsinki, Gustaf H\"allstr\"omin katu 2, FI-00014 Helsinki, Finland}

\author{Thorsten Naab}
\affiliation{Max Planck Institute for Astrophysics, Karl-Schwarzschild-Str. 1, D-85740, Garching, Germany}

\author{Peter H. Johansson}
\affiliation{Department of Physics, University of Helsinki, Gustaf H\"allstr\"omin katu 2, FI-00014 Helsinki, Finland}

\author{Bruce Elmegreen}
\affiliation{IBM T. J. Watson Research Center, 1101 Kitchawan Road, Yorktown Heights, NY 10598, USA}

\author{Chia-Yu Hu}
\affiliation{Max Planck Institute for Extraterrestrial Physics, Giessenbachstrasse 1, D-85748, Garching, Germany}

\author{Stefanie Walch}
\affiliation{I. Physikalisches Institut, Universit\"at zu K\"oln, Z\"ulpicher Strasse 77, D-50937 K\"oln, Germany}
\affiliation{Cologne Center for Data and Simulation Science, University of Cologne, www.cds.uni-koeln.de}





\begin{abstract}
We analyze the three-dimensional shapes and kinematics of the young star cluster population forming in a high-resolution \griffin\ project simulation of a metal-poor dwarf galaxy starburst. The star clusters, which follow a power-law mass
distribution, form from the cold ISM phase with an IMF sampled with 
individual stars down to 4 solar masses at sub-parsec spatial resolution. Massive stars and their important feedback mechanisms are modelled in detail. 
The simulated clusters follow a surprisingly tight relation between the specific angular momentum and mass with indications of two sub-populations. Massive clusters ($M_\mathrm{cl}\gtrsim 3\times 10^4 M_{\odot})$ have the highest specific angular momenta at low ellipticities ($\epsilon\sim 0.2$) and show alignment between their shapes and rotation. 
Lower mass clusters have lower specific angular momenta with larger scatter, show a broader range of elongations, and are typically misaligned indicating that they are not shaped by rotation. 
The most massive clusters $(M \gtrsim 10^5\,M_{\odot})$ accrete gas and proto-clusters from a $ \lesssim 100\,\rm pc$ scale local galactic environment on a  $t \lesssim 10\,\rm Myr$ timescale, inheriting the ambient angular momentum properties. Their two-dimensional kinematic maps show ordered rotation at formation, up to $v \sim 8.5\,\rm km s^{-1}$, consistent with observed young massive clusters and old globular clusters, which they might evolve into. The massive clusters have angular momentum parameters $\lambda_R\lesssim 0.5$ and show  Gauss-Hermite coefficients $h_3$ that are anti-correlated with the velocity, indicating asymmetric line-of-sight velocity distributions as a signature of a dissipative formation process.

\end{abstract}



\section{Introduction}

The oldest and densest self-gravitating stellar systems, globular clusters (GCs), were long considered to be non-rotating, spherically symmetric dense objects. 
However contrary to this notion, rotational motion at least at a level of a few \mbox{km s$^{-1}$} seems to be a common feature, as has been recently observed in many of the GCs in the Milky Way \citep[e.g.][]{2012A&A...538A..18B, 2014ApJ...787L..26F}. 

Currently, it is also possible to survey the velocity distributions in dozens of Milky Way GCs using IFU spectroscopy and direct astrometry  \citep{2018ApJ...860...50F, 2018MNRAS.473.5591K} with a most recent major contribution by the Gaia survey (e.g. \citealt{2015A&A...573A.115L,2018MNRAS.481.2125B}). 
The radial rotational profiles typically span from a few \mbox{km s$^{-1}$} to slightly above $ 10$ \mbox{km s$^{-1}$}, whereas the peak velocity to central velocity dispersion ($V/\sigma$) ratios are typically below unity ($0 \lesssim V/\sigma \lesssim 0.5$ in the Gaia survey data). Observations of local young massive clusters (YMCs), which are the potential present-day analogues for forming globular clusters \citep{2010ARA&A..48..431P,2014prpl.conf..291L}, also show systemic rotation \citep{2012A&A...546A..73H}, as do some intermediate age massive stellar clusters \citep{2013ApJ...762...65M,2018MNRAS.480.1689K,2019MNRAS.483.2197K}. This indicates that GC rotation might have been generated at their formation. On the other hand, lower mass star clusters are in general dispersion dominated, showing low levels of ordered rotation \citep{2019ApJ...870...32K}. 

One of the commonly used indicators for rotation in studies of dynamically hot stellar systems is the shape deformation. In very massive objects, such as elliptical galaxies, the ellipticity tends to increase with increasing rotational support \citep{2011MNRAS.414..888E}. Contrary to galaxies, there does not seem to be a clear correlation between rotation and elongation for stellar clusters. For example, open stellar clusters in the Milky Way have ellipticities in the range from 0 to $\sim0.4$ with an observed peak at $\sim0.2$ \citep{2009A&A...495..807K} while old massive GCs have ellipticities up to $\epsilon \sim 0.3$, concentrating around lower values (sample mean of the order $\epsilon \lesssim 0.1$, \citealt{2010arXiv1012.3224H,2010ApJ...721.1790C}). Compared to the notion that massive star clusters tend to rotate while lower mass clusters do not, the elongated elliptic shapes of lower mass star clusters has been attributed to the galactic tidal field instead of internal rotation \citep{2012MNRAS.426.2427S}. 

The stellar mass and internal velocity structure in old star clusters may hold information about their formation histories \citep{2010A&A...516A..55C,2015MNRAS.454.4197R}. For example, differences in the velocity distributions of the chemically or age-wise differentiated multiple populations (\citealt{2008MNRAS.390..693D,2012A&ARv..20...50G}, see \citealt{2020SSRv..216...69A} for a recent review) observed in Galactic \citep{2017MNRAS.464.3636M} and Magellanic Cloud \citep{2008ApJ...681L..17M,2020MNRAS.491..515M} GCs could be used to assess whether there were differences in the formation mechanisms of the first and second generation stars \citep{2013MNRAS.429.1913V}. 
The second generation stars tend in general to show more rotational support and more radially biased velocity distributions \citep{2020ApJ...889...18C}, as is also seen in simulated clusters \citep{2019A&A...622A..53B}. The low amount of rotation seen in the first generation could for example be attributed to protocluster mergers \citep{2019A&A...622A..53B}. The formation conditions could thus be imprinted e.g. in the velocity anisotropy profiles of the populations \citep{2018MNRAS.479.5005M,2019MNRAS.487.5535T,2020ApJ...889...18C}. The specific dynamical evolution of GCs however is a complex combination of both the initial velocity structure and the surrounding tidal field \citep{2016MNRAS.455.3693T}. 

Star formation which results in star clusters takes several free-fall times (see e.g. a review by \citealt{2019ARA&A..57..227K}), and is therefore a prolonged process which requires fueling of gas which is ultimately converted into stars. Initial rotation and possible asymmetries in the fragmenting giant molecular clouds, as well as the galactic tidal field in the cluster formation environment, may result in a net non-zero rotation in the star-forming gas, as angular momentum is conserved. The angular momentum may consequently be translated even down to stellar scales in the form of aligned stellar spins and binary star orbits \citep{2017NatAs...1E..64C,2019MNRAS.483.2197K}. 

Rotation in massive star clusters affects their evolution, as net angular momentum may speed up the core collapse and increase the escape rate of stars \citep{1999MNRAS.302...81E,2007MNRAS.377..465E,2013MNRAS.430.2960H,2020arXiv200600004C}. Hydrodynamical studies at individual cloud-scales, even without external dynamical effects such as a surrounding live galaxy, find it extremely difficult to produce non-rotating clusters \citep{2016A&A...591A..30L, 2020MNRAS.496...49B} with only very low-mass (e.g. $<100\, M_\odot$, \citealt{2017MNRAS.467.3255M}) clusters exhibiting no clear rotational signatures. On the other hand, because stellar mass is lost to the surrounding tidal field and angular momentum is re-distributed outwards due to two-body relaxation on a relaxation time scale, rotation of star clusters is expected to decrease during their lifetimes \citep{1999MNRAS.302...81E,2007MNRAS.377..465E,2008MNRAS.383....2K,2012MNRAS.425.2872H,2017MNRAS.469..683T}. Thus any signs of rotation in objects older than 10 Gyr indicate more rotational support closer to their formation. 

Direct $N$-body simulations without hydrodynamics and the galactic environment represented by a point mass tidal field (e.g. \citealt{2017MNRAS.469..683T}) have recently been used to investigate the velocity and angular momentum properties in evolving star clusters. Similar investigations have also been carried out using 
isolated cloud-scale simulations with hydrodynamics but without the global galactic environment (e.g. \citealt{2017MNRAS.467.3255M}). \citet{2019A&A...622A..53B} also studied globular clusters forming in a gas-rich dwarf galaxy, with special emphasis on their multiple stellar populations, but only briefly concluded that they see rotation in the clusters. 

In this paper we study in detail the angular momentum properties of the full population of star clusters formed in a simulated gas-rich dwarf galaxy starburst within the \griffin\ project introduced in \citeauthor{2020ApJ...891....2L} (2020, L20 hereafter). Dwarf starbusts are potentially good analogues of the turbulent star formation processes prevalent at high redshifts and can therefore be used to probe the conditions where present-day globular clusters might have formed. Here  we pay special attention to the most massive star clusters which show globular cluster-like properties as they exceed 100 Myr in mean stellar age. The formation of these three  massive clusters was already discussed in \citeauthor{2019ApJ...879L..18L} (2019, L19 hereafter). We study in particular how the velocity profiles of the clusters are built and demonstrate that the clusters inherit the angular momentum of their progenitor gas clouds. We follow the evolution of the angular momentum as the clusters evolve in the global galactic environment, and compare their rotational profiles to globular clusters such as observed in the Gaia DR2 \citep{2018MNRAS.481.2125B} and with the MUSE instrument \citep{2018MNRAS.473.5591K}. The accretion of globular clusters formed in dwarf galaxies by more massive galaxies has been proposed as the origin of many, if not the majority of the Milky Way globulars \citep{2019A&A...630L...4M}. Here we study a simulated dwarf starburst, as a direct progenitor of such globular clusters. 

The paper is organized as follows. In Section 2 we describe the numerical methods, the initial simulation setup and the analysis tools used in this study. In Section 3 we review the kinematic and shape properties of the young star cluster population forming during the dwarf starburst. Section 4 discusses the formation and evolution of the three most massive clusters in our simulation sample. In Section 5 we construct two dimensional kinematic maps of the most massive clusters, and investigate their projected rotational properties and compare to recent observations of globular clusters rotation. Finally, we present our conclusions in Section 6. 
 
\section{Simulations and analysis}

\subsection{Simulation code}

The simulation data analysed in this paper is a subset of the output from the gas-rich dwarf galaxy merger simulations that are a part of the \griffin\ project, introduced in L20. The simulations, described in detail in L19 and L20, were ran at $\sim 4 \, M_\odot$ baryonic mass resolution and with $0.1\,\rm pc$ baryonic gravitational softening length using \sphgal\ (\citealt{2014MNRAS.443.1173H, 2016MNRAS.458.3528H, 2017MNRAS.471.2151H}), an updated version of the tree/smoothed particle hydrodynamics (SPH) code  \gadget\ \citep{2005MNRAS.364.1105S}. This code version uses the pressure -- energy formulation of the hydrodynamic equations, and a Wendland $\rm C^4$ smoothing kernel over the $100$ nearest neighbouring gas particles. Artificial viscosity and conduction of thermal energy are modelled as detailed in \citet{2014MNRAS.443.1173H}.


The non-equilibrium chemical network, adapted from \citet{2007ApJS..169..239G}, enables the modelling of the evolution of the ISM down to temperatures of \mbox{$T=10$ K}. At low temperatures we follow six chemical species (H$_{2}$, H$^{+}$, H, CO, C$^{+}$, O) and the free electron density, and assume a constant dust-to-gas mass ratio of $0.1\%$. At high temperatures, above $T=3\times10^4$ K, the gas cooling follows the metallicity-dependent equilibrium rates from \citet{2009MNRAS.393...99W}. The optically thin high-temperature gas is also assumed to be in ionization equilibrium with a cosmic UV-background field from \citet{1996ApJ...461...20H}.


Gas in a converging flow is allowed to form stars at a $\epsilon_\mathrm{SF}=2\%$ statistical efficiency if the locally averaged quantities resolve the Jeans mass with 8 SPH kernel masses. That is, the probability for a gas particle to be turned into stars is $1-\exp(-p)$ where $p=\epsilon_\mathrm{SF}\Delta t/t_\mathrm{ff}$,
and $\Delta t$ and $t_\mathrm{ff}$ are the time step length and free-fall time, respectively. 
In addition, we enforce star formation at high densities, when the local Jeans mass drops below half a kernel mass and turn such gas particles into stars instantaneously.


The mass of each newly formed star particle is sampled and stored as an array of stellar masses along the Kroupa IMF. Gas particles which draw a stellar mass from $4\, M_\odot$ upwards are realised into single massive stars, while smaller masses are sampled into stellar masses until the total mass comprises the mass of the original gas particle. If the final mass of the star particle exceeds the mass of the progenitor gas particle, the excess mass is acquired from nearby dense gas particles thus conserving the total mass. 


The stellar feedback models include hydrogen-ionising radiation, the FUV radiation field assuming optically thin gas \citep[e.g.][]{2017MNRAS.471.2151H}, energy and ejecta from type II supernovae (SNII, \citealt{2004ApJ...608..405C}) and winds from AGB stars \citep{2010MNRAS.403.1413K}. The FUV radiation creates a temporally and spatially evolving interstellar radiation field, which is propagated into the ISM along 12 lines of sight, while accounting for dust extinction. With our mass resolution the momentum and hot phase generation by individual SNII events releasing $10^{51}$ erg into the surrounding SPH kernel region is resolved at the typical ambient densities (see e.g. \citealt{2020MNRAS.495.1035S} and L20). This is particularly important for capturing SN driven outflows \citep[see e.g.][for a discussion]{2017ARA&A..55...59N}.

\begin{figure*}
\includegraphics[width=\textwidth]{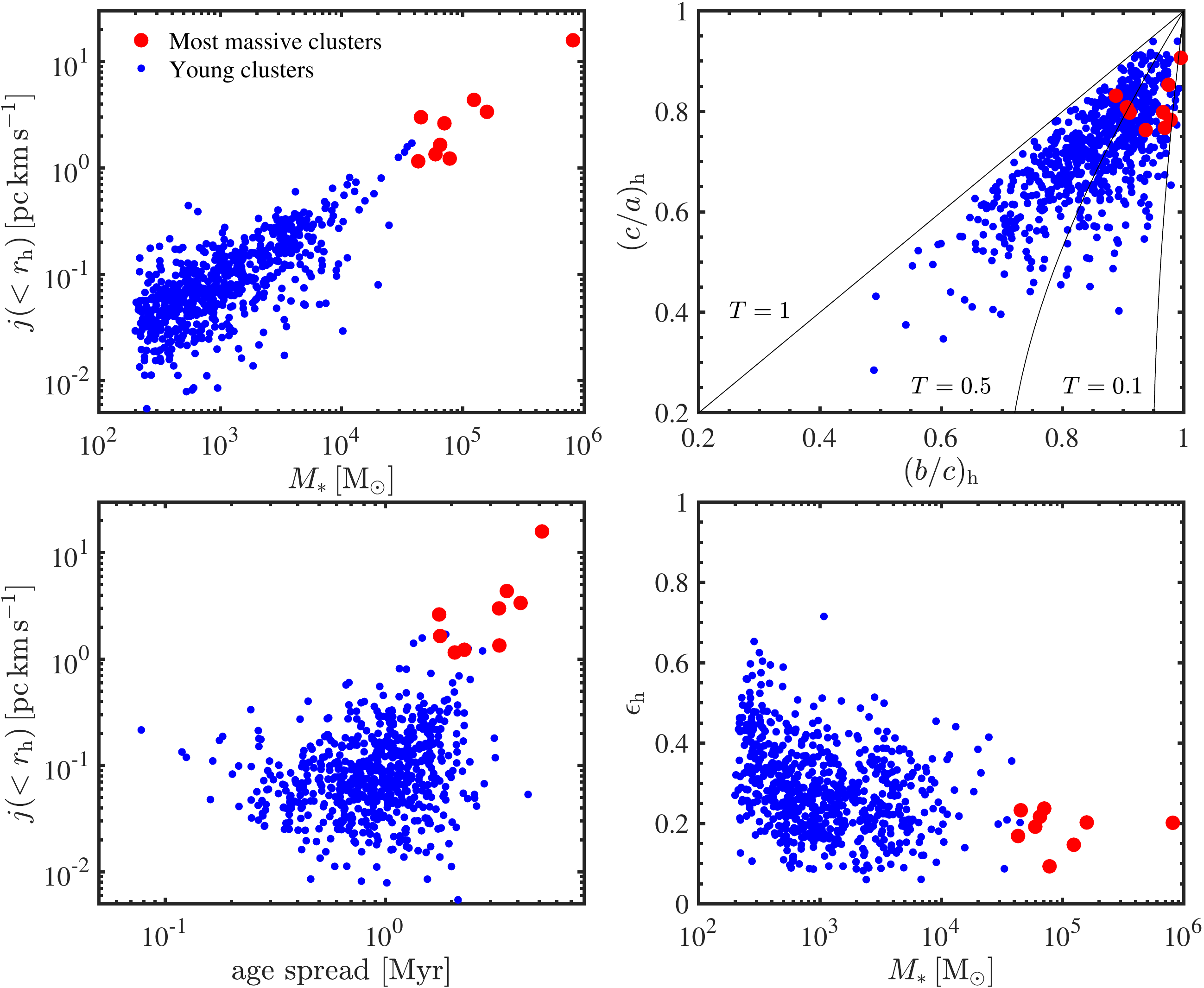}
\caption{The angular momentum and shape of young stellar clusters calculated within the half-mass radius. On the left we
show the specific angular momentum as a function of cluster mass (top) and stellar age spread (bottom). 
In the right column we show the axis ratios of the three principal axes $a$, $b$ and $c$ (top), and cluster
ellipticities as a function of mass (bottom). The clusters are analysed immediately after the starburst at t=175 Myr. The lines in the top right panel indicate constant triaxiality $T$ (see text for details). The data shown includes only young clusters with mean stellar ages below 20 Myr and the red points highlight the nine most massive clusters which exhibit signs of rotation, as discussed in Section \ref{section:los_maps}. \label{fig:specific_j}}
\end{figure*}

\subsection{Initial conditions}

The initial conditions have previously been introduced in detail in \citet{2016MNRAS.458.3528H} and L20. Briefly, we set up two identical gas-rich dwarf galaxies with virial masses of $M_\mathrm{vir}=2\times 10^{10}\, M_\odot$. Both dwarfs consist of a dark matter halo with a Hernquist density profile and an exponential stellar disk, with a small baryon mass fraction of $0.3\%$ and a high disk gas mass fraction of $66\%$. The dark matter mass resolution is set to $10^4\, M_\odot$ per particle. The gaseous and stellar disks have total masses of $M_\mathrm{gas}=4\times10^7\, M_\odot$ and $M_\mathrm{*}=2\times10^7\, M_\odot$, resolved initially with $4\, M_\odot$ per particle. The stellar disks have scale radii of $0.73\,\rm kpc$ and scale heights of $0.35\,\rm kpc$, while the more extended gas disks have scale radii of $1.46\,\rm kpc$ and scale heights calculated such as to set them in hydrostatic equilibrium \citep{2005MNRAS.361..776S}. 

The softening length of the gravitational force in the simulation is set to $63\,\rm pc$ for dark matter particles and $0.1\, \rm pc$ for baryonic particles. We set the two galaxies on parabolic orbits with initial and pericentric distances of $r_\mathrm{init}=5\,\rm kpc$ and $r_\mathrm{peri}=1.46\,\rm kpc$, where the galactic disks have been oriented off the orbital plane with inclinations and the arguments of pericenter set as $\{i_{1}, i_{2}\}=\{60^{\circ}, 60^{\circ}\}$ and $\{\omega_{1}, \omega_{2}\}=\{30^{\circ}, 60^{\circ}\}$.

\subsection{Identification of young star clusters}

We search for bound star clusters in the simulation snapshots using the friends-of-friends and \subfind\ algorithms in \gadget\ \citep{2001MNRAS.328..726S, 2009MNRAS.399..497D}. Any bound stellar object with more than 50 stellar particles
($\sim 200\,M_\odot$) after the \subfind\ routine has executed the unbinding procedure 
is considered a star cluster. Here we concentrate mostly on the most massive clusters, which can easily be  followed individually during their formation and subsequent evolution once their constituent particles have been identified. 

\subsection{Intrinsic shape and angular momentum}

We calculate the intrinsic three-dimensional shape of the star clusters using the moment of inertia. The directions of the principal axes  $a$, $b$ and $c$ and their axis ratios are obtained from the eigenvectors and eigenvalues of the moment of inertia tensor. The ellipticity is then defined using the ratio between the longest and shortest principal axes (i.e. semi-major and semi-minor axes) as $1-c/a$.
We also calculate the direction and components of the angular momentum $L_x$, $L_y$, and $L_z$ in 
both the stellar and gaseous particles along the formation sequence with respect to the center of mass of all gas and star particles which will end up bound in each cluster. We produce cumulative radial angular momentum profiles \mbox{$L(<r)=\sqrt{L_x^2(<r)+L_y^2(<r)+L_z^2(<r)}$} and specific angular momentum profiles \mbox{$j(<r)=L(<r)/m(<r)$}, as well as the total value of $j$ for particles within the half-mass radius, $j(<r_\mathrm{h})\equiv j_\mathrm{h}$. To investigate trends in the full young cluster population we show the characteristic shape and angular momentum properties of the population in Fig. \ref{fig:specific_j} and discuss the results in Section \ref{Sec_3}. 

\subsection{Two-dimensional kinematic maps}\label{sec:kinematics}

We study the projected line-of-sight velocity distribution (LOSVD) in the massive clusters following the methods often used in IFU studies and in stellar astrometry \citep{2007MNRAS.379..418C,2011MNRAS.414.2923K}. Such methods typically include tesselation of the underlying data (regular pixel-averaged data or single stars) to obtain constant signal-to-noise maps such as presented e.g. in \citet{2006MNRAS.366.1126C} for early-type galaxies and in \citet{2018MNRAS.481.2125B} for globular clusters. The same methods can also essentially be used in analysing numerical simulations \citep{2014MNRAS.444.3357N,2018MNRAS.475.3934L,2018ApJ...864..113R,2019MNRAS.489.2702F}. As we are modeling individual massive stars rather than massive stellar population particles, we discard seeing effects applied to larger scale simulations \citep{2014MNRAS.444.3357N}, and instead directly fit the LOSVD to the Voronoi-tesselated particle data. 

First we orient the clusters edge-on according to the direction of the angular momentum vector, as described in Section \ref{section:angular_vector}. We remove the effect of inclination on the projected velocity data and thus study the most extreme systemic rotation in each cluster.
Next, the particles in each cluster are projected onto a 2D grid with a pixel resolution of approximately 0.1 pc following the gravitational softening length. A Voronoi tessellation scheme \citep{2003MNRAS.342..345C} is used to divide the underlying regular pixels into Voronoi cells with roughly equal number of particles in each cell, ensuring at least 100 stars per cell, following the data reduction of \citet{2018MNRAS.481.2125B}. In the most massive cluster where the central region includes more than $10^5$ particles, we additionally limit the number of Voronoi cells to 500. The mean velocity and velocity dispersion are then calculated from the particles in each Voronoi spaxel either through the mass weighted mean and standard deviation or by fitting Gauss-Hermite function \citep{1993ApJ...407..525V} which provides the higher order deviations ($h_3$ and $h_4$ reminiscent of the skewness and kurtosis) from a Gaussian. 

We restrict the analysis to the central regions of the clusters (i.e. within a couple half-mass radii) thus matching the typical spatial extent of the current observational surveys \citep{2018MNRAS.473.5591K}. The number of spaxels spans thus from a few hundred down to $\sim40$ in the smallest mass clusters for which we still are able to produce spatially resolved results. Using this analysis technique we are able to resolve the nine 
most massive clusters, with the limiting factor being mainly the number of stars in the relevant analysis regions. 

\section{Shape and kinematics of the star cluster population}
\label{Sec_3}

\begin{figure*}
\includegraphics[width=\textwidth]{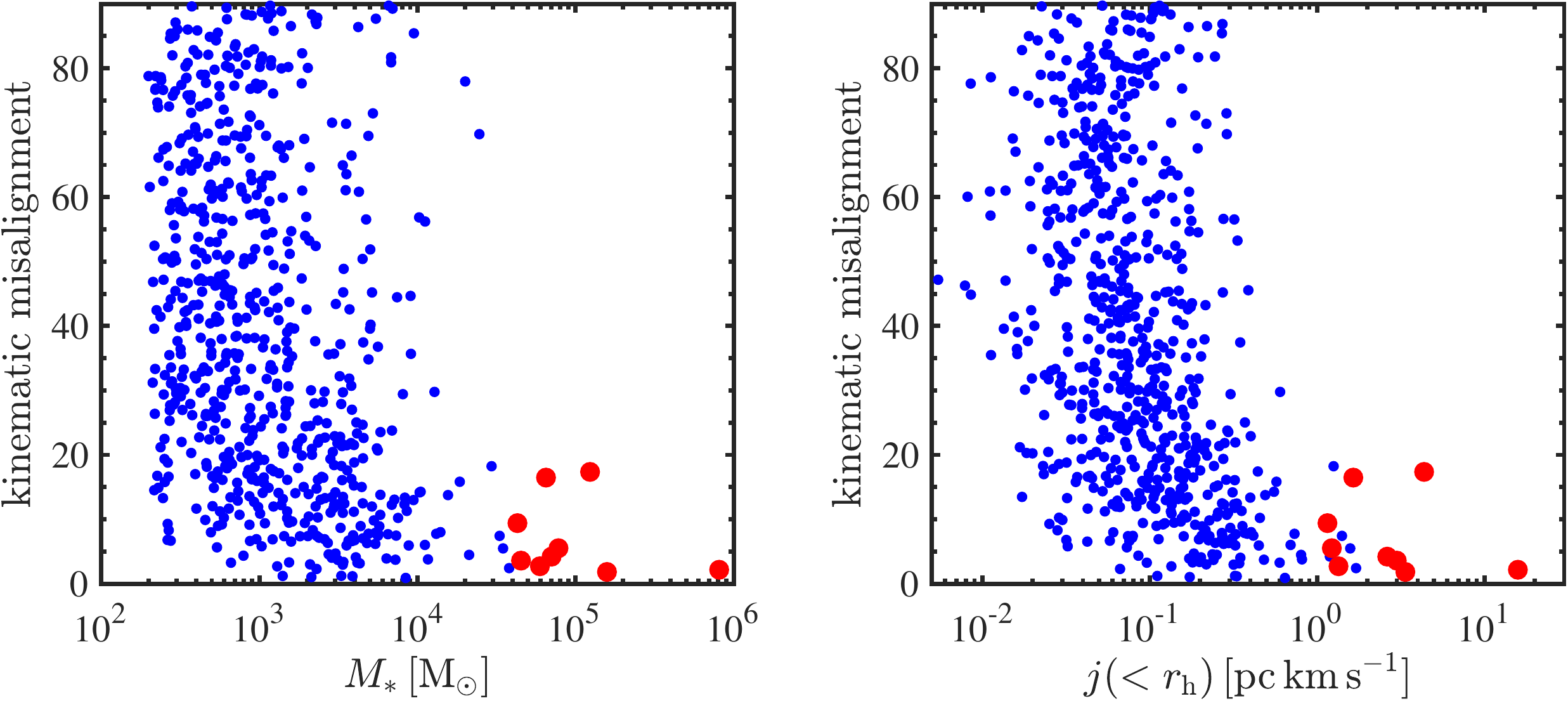}
\caption{The angular misalignment between the moment of inertia (shortest principal axis) and the angular momentum vector in degrees within the half-mass radius in the young ($<20$ Myr old) star clusters as a function of stellar mass (left) and the specific angular momentum (right) immediately after the starburst. The red points highlight the nine massive clusters we investigate in more detail. \label{fig:misalignment}}
\end{figure*}

\subsection{Kinematic analysis of the cluster population}

In the top left panel in Fig. \ref{fig:specific_j} we show the specific angular momentum $j_\mathrm{h}$ as a function of the star cluster mass immediately after the starburst at $t=175 \ \rm Myr$. We show here young star clusters with mean stellar ages below $20$ Myr, highlighting the results for the nine massive clusters for which we have been able to produce kinematic maps (see e.g. Section \ref{section:los_maps}). These nine clusters have mean ages between 1 and 18.6 Myr. 
 
There is a clear trend for higher mass clusters to have higher specific angular momenta. A power-law fit to the data in the top left panel of Fig. \ref{fig:specific_j} reveals a power-law with $j_\mathrm{h}\propto M^{0.62}$ when fit to all data, \mbox{$j_\mathrm{h}\propto M^{0.71}$} when only clusters above stellar mass $10^3\,M_\odot$ are considered and $j_\mathrm{h}\propto M^{0.97}$ when fit to only $>10^4\,M_\odot$ clusters. 

Assuming the clusters form from clouds with $\rho \sim 1/r$ density profiles the enclosed mass scales as $M(r) \sim r^2$ and gravity ($v^2=G M(r)/r$) introduces a rotation speed of $v(r) \sim \sqrt{r}$ with a resulting specific angular momentum of 
\begin{equation}
j(r) \sim r v \sim r^{3/2} \sim M(r)^{3/4}. 
\end{equation}

With an average density scaling as $\sim 1/\sqrt{M}$ for each cluster forming region the rotation speed scales as $v \sim M^{1/4}$ with an average size $r \sim \sqrt{M}$. This results in 
\begin{equation}
j \sim M^{3/4}
\end{equation}
Such a scaling is therefore consistent with the Larson relations ($M \sim r^2; v \sim \sqrt{r}$, \citealt{1981MNRAS.194..809L}) for gravitating clouds. In a separate study (Fotopoulou et al., in preparation) we will show that the star forming clouds in this simulation follow the Larson relations well. For steeper cloud profiles, i.e. $\rho \sim r^{-2}$, the specific angular momentum would scale linearly with mass. \citet{2020arXiv200600004C} have used three idealised star forming cloud setups with varying density profiles and find more massive stars clusters with higher specific angular momentum for steeper initial cloud density profiles.

Clusters with higher stellar mass collect their material from a larger region (as will be discussed e.g. in Fig. \ref{fig:init_gas_radii}), thus ending up inevitably with higher values of angular momentum. A similar trend is seen when $j_\mathrm{h}$ is compared to the age spread in the clusters in the bottom left panel of Fig. \ref{fig:specific_j}. Higher mass clusters are built from larger sized regions on longer time scales, resulting in larger age spreads. 

Next, we investigate the shapes of the young star cluster population in the right hand panels of Fig. \ref{fig:specific_j} by showing the axial ratios $c/a$ and $b/a$ and ellipticities $\epsilon=1-c/a$ within half-mass radii in the young clusters. The lines in the top right panel denote constant values of triaxiality $T=(a^2-b^2)/(a^2-c^2)$ at $T=1$ (prolate), $T=0.5$ (triaxial) and $T=0.1$ (close to oblate). The massive clusters are mostly more spherical and have oblate shapes compared to the general cluster population, which is also seen in the ellipticity. 

Young star clusters tend to form in hydrodynamical simulations with non-zero ellipticities \citep{2017MNRAS.467.3255M} and we see a trend for increasing ellipticity with decreasing cluster mass. In observations, globular clusters tend to be less elliptical compared to for example open clusters, and we see a similar trend even though the less massive clusters have less specific angular momentum compared to the massive clusters. $N$-body simulations \citep{2013MNRAS.430.2960H} and analytical studies \citep{2006MNRAS.373..677F} have shown that ellipticities tend to decrease as the clusters evolve. Therefore it is not too surprising that we see more non-spherical shapes when comparing our massive clusters to old observed clusters which have on average $\epsilon<0.1$ \citep{2010arXiv1012.3224H}. We also note that we use the most extreme axis ratio ($1-c/a$) and do not consider projection effects, which have a decreasing effect (see Section \ref{section:los_maps}) on the observed ellipticities.

We then calculate the angle between the moment of inertia (i.e. direction of the shortest principal axis) and direction of the angular momentum vector, and show the results as a function of stellar mass and specific angular momentum in Fig. \ref{fig:misalignment}. The shortest principal axis of the massive clusters tends to be aligned with the rotational axis while lower mass clusters do not show clear correlation between their shapes and rotation. The highest angular momentum clusters are also the least misaligned while the clusters with less angular momentum can be either aligned or very misaligned. Combined with the shape and angular momentum information derived in Fig. \ref{fig:specific_j}, we conclude that the non-spherical shape of the lower mass clusters is probably not due to rotation as is the case e.g. in more massive stellar objects such as rotating elliptical galaxies \citep{2011MNRAS.414..888E}. 

\begin{figure}
\includegraphics[width=\columnwidth]{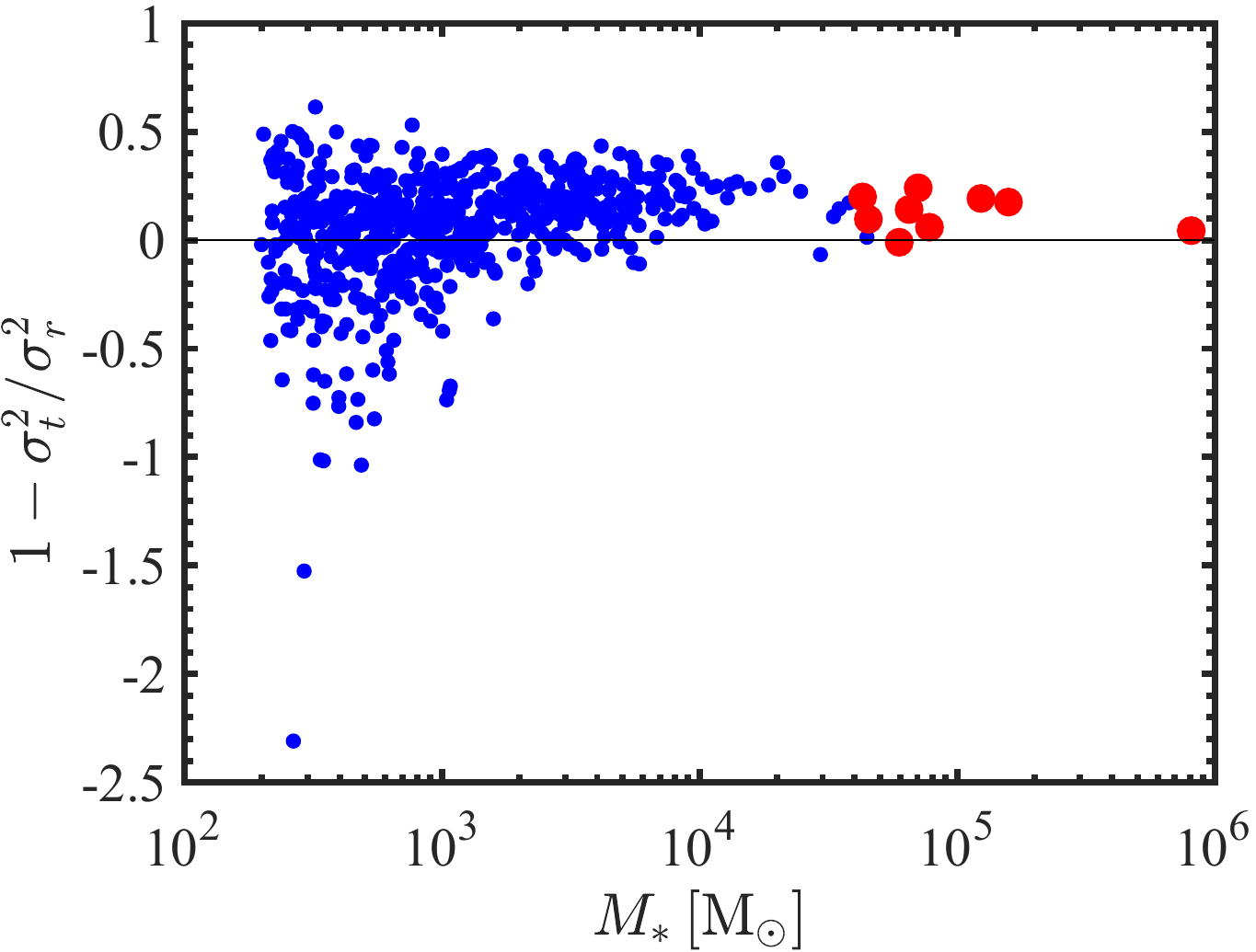}
\caption{The velocity anisotropy within the half-mass radius in star clusters with mean stellar ages less than 20 Myr as a function of their bound stellar mass immediately after the starburst. The red points highlight the nine massive clusters we investigate in more detail. \label{fig:dispersion}}
\end{figure}

\begin{figure}
\includegraphics[width=\columnwidth]{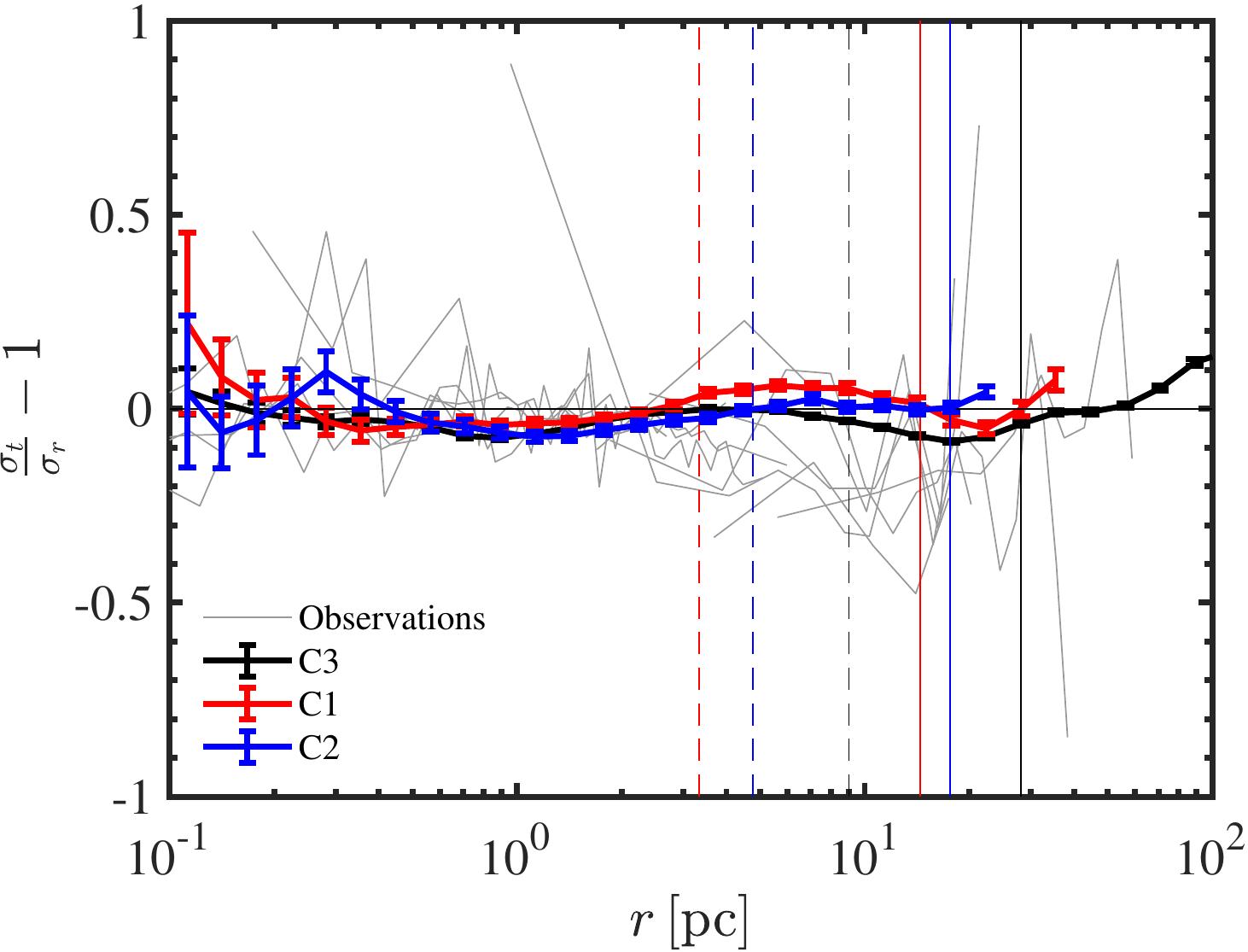}
\caption{The velocity anisotropy in the three most massive clusters 100 Myr after their formation, compared to observed anisotropies from \citet{2015ApJ...803...29W} and \citet{2019MNRAS.487.3693J} for globular clusters which have been identified with a clear rotational signal in \citet{2018MNRAS.481.2125B}. The vertical lines show the half-mass radii and $r_{80}$ of the chronologically labeled clusters C1--C3, see Sections \ref{sec:assembly} and \ref{sec:formation} for details. \label{fig:beta}}
\end{figure}

\begin{figure*}
\includegraphics[width=0.98\textwidth]{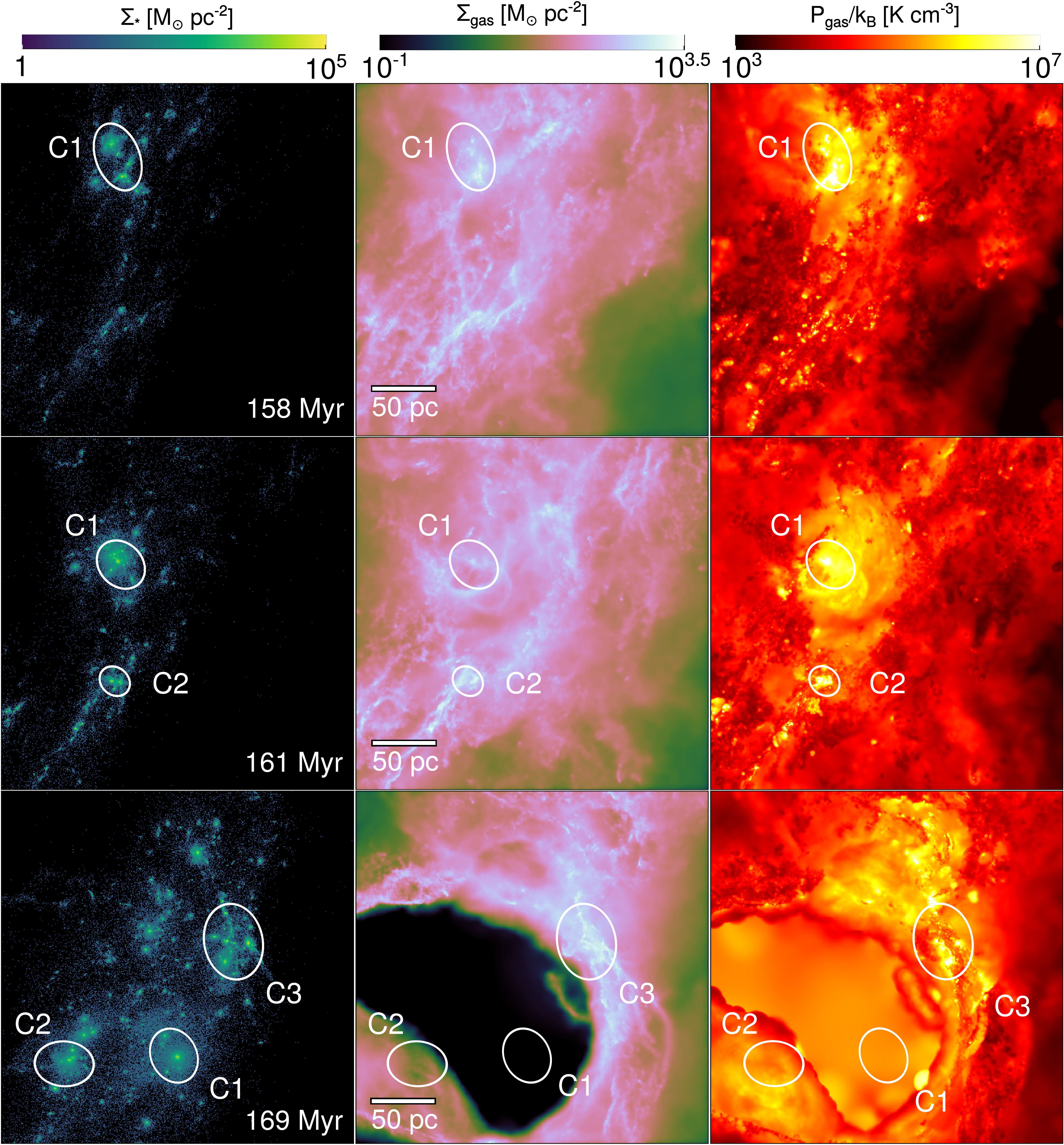}
\caption{The surface density of stars formed during the simulation (left column), the surface density of the gas (middle column) and the corresponding gas thermal pressure (right column) are shown in a projected 300 pc by 300 pc by 300 pc  cube showing the birth sites of the three most massive stellar clusters (white ellipses). Each row highlights one of the clusters and the gaseous environment at a time when half of the final cluster stellar mass has been formed, in chronological order from top to bottom (C1, C2, C3). The white ellipses illustrate the regions from which most of the mass will end up in the corresponding cluster. The most massive cluster (C3) forms in a dense filament swept up by a super shell generated by supernovae exploding in C1. \label{3_clusters}}
\end{figure*}

\subsection{Velocity anisotropy}

The velocity anisotropy is often used to quantify the kinematic structure of stellar systems. The exact definition of velocity anisotropy varies from study to study, and here we denote it using the ratio between the tangential velocity dispersion $\sigma_t=\sqrt{(\sigma_\theta^2+\sigma_\phi^2)/2}$, and the radial velocity dispersion $\sigma_r$ as $1-\sigma_t^2/\sigma_r^2$. The system is isotropic when all three components are equal. We show the anisotropy in the young clusters within the half-mass radius in Fig. \ref{fig:dispersion} as a function of the stellar mass. The clusters show radial anisotropy as a relic of their formation process. However, the less massive clusters show a larger spread towards both radial and tangential anisotropy. If we look at the velocity dispersions in cartesian directions along the principal axes of the clusters, all of the massive clusters show anisotropy along the plane in which they are elongated (perpendicular to the shortest principal axis). The lower mass clusters on the other hand span again a wide range from perpendicular to parallel anisotropy with respect to their shape. 

In Section \ref{sec:formation} we separate the bound stars into in-situ formed and accreted stars in the three most massive clusters according to their formation radius from the local center of mass. The anisotropy of the accreted stars is $\sim0.7$ in the three clusters while the in-situ formed stars have values between $-0.07$ and $0.15$. The accreted stars are therefore very radially biased while the in-situ formed stars have fairly isotropic velocity distributions.

As the simulated clusters evolve, the mean anisotropy within $r_\mathrm{h}$ depicted in Fig. \ref{fig:dispersion} in the entire population tends to move towards isotropy with increasing age. The relaxation time scale within the half-mass radius in the most massive cluster when it reaches the highest central density ($t_\mathrm{sim}\sim171$ Myr, see L19) is of the order of 140 Myr. At our mass resolution with mean to maximum stellar mass ratio of $\sim0.1$ this corresponds to a mass segregation time scale of $\sim 14$ Myr \citep{1969ApJ...158L.139S, 2002ApJ...570..171F} and an even shorter time scale in lower mass clusters.
We next take a look at the radial velocity anisotropy in the three most massive clusters after a few segregation time scales when they reach 100 Myr in mean stellar age. These clusters have been shown in L19 to be good GC candidates already at an age of 100 Myr. Here we use another definition for the radial versus tangential velocity anisotropy from the observational literature \mbox{$\sigma_t/\sigma_r -1$}. The tangential dispersion is again defined as $\sigma_t=\sqrt{(\sigma_\phi^2+\sigma_\theta^2)/2}$ as in Fig. \ref{fig:dispersion} in order to set pure isotropy at zero. Note however that the radially and tangentially biased directions in the anisotropy are reversed to the ones used in Fig. \ref{fig:dispersion}. 

The results are shown for the three 
100 Myr old clusters in Fig. \ref{fig:beta}. The errorbars have been obtained by bootstrapping the particles 1000 times. 
We compare the resulting anisotropy profiles to six observed globular clusters with clear rotational signal (NGC 104, NGC 5139, NGC 5904, NGC 6656 and NGC 6752) identified in \citet{2018MNRAS.481.2125B} and obtained from \citet{2015ApJ...803...29W} (data at inner radii) and \citet{2019MNRAS.487.3693J} (data at outer radii).

The 100 Myr old massive clusters show fairly isotropic velocity distributions in Fig. \ref{fig:beta}, with still a slight tendency for radial anisotropy as already indicated at younger ages in Fig. \ref{fig:dispersion}. The observations show a very similar picture, with a slight tendency for more tangential values (consistent with zero anisotropy within errors) in the central regions which transitions to slight radial anisotropy in the outer regions. However, note that we compare here the intrinsic velocity dispersions of the simulated 100 Myr old clusters with the projected tangential and radial dispersions of very old ($>10$ Gyr) clusters. We find that despite of this the simulated profiles show very low anisotropy, in good agreement with the observations.

\begin{figure}
\includegraphics[width=\columnwidth]{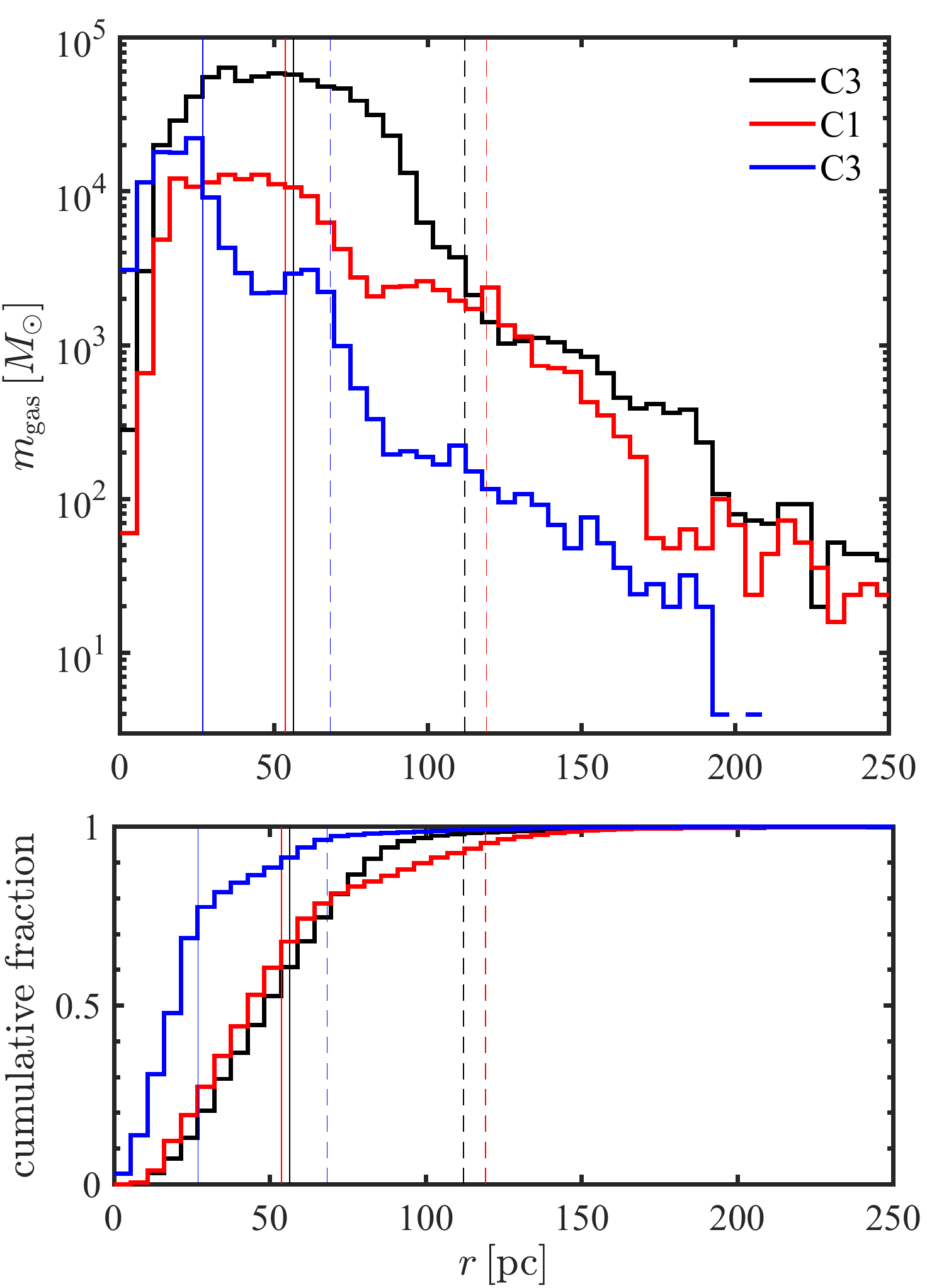}
\caption{Top: the initial radii of the gas particles which will constitute the stellar mass of the clusters C1--C3 at a cluster age of 10 Myr. Bottom: the cumulative mass fraction calculated from the top panel. The radii are measured in the last snapshots respectively where only a couple per cent of the final bound mass has formed, i.e. at the start of the star formation of the final mass in each three clusters. The distances have been calculated with respect to the center of mass of all gas particles which end up in the clusters. The vertical lines show the mean (solid) and mean+$2\sigma$ (dashed) values. The majority of the cluster material is assembled from a region within $\sim 100$ pc. \label{fig:init_gas_radii}}
\end{figure}

\begin{figure*}
\includegraphics[width=\textwidth]{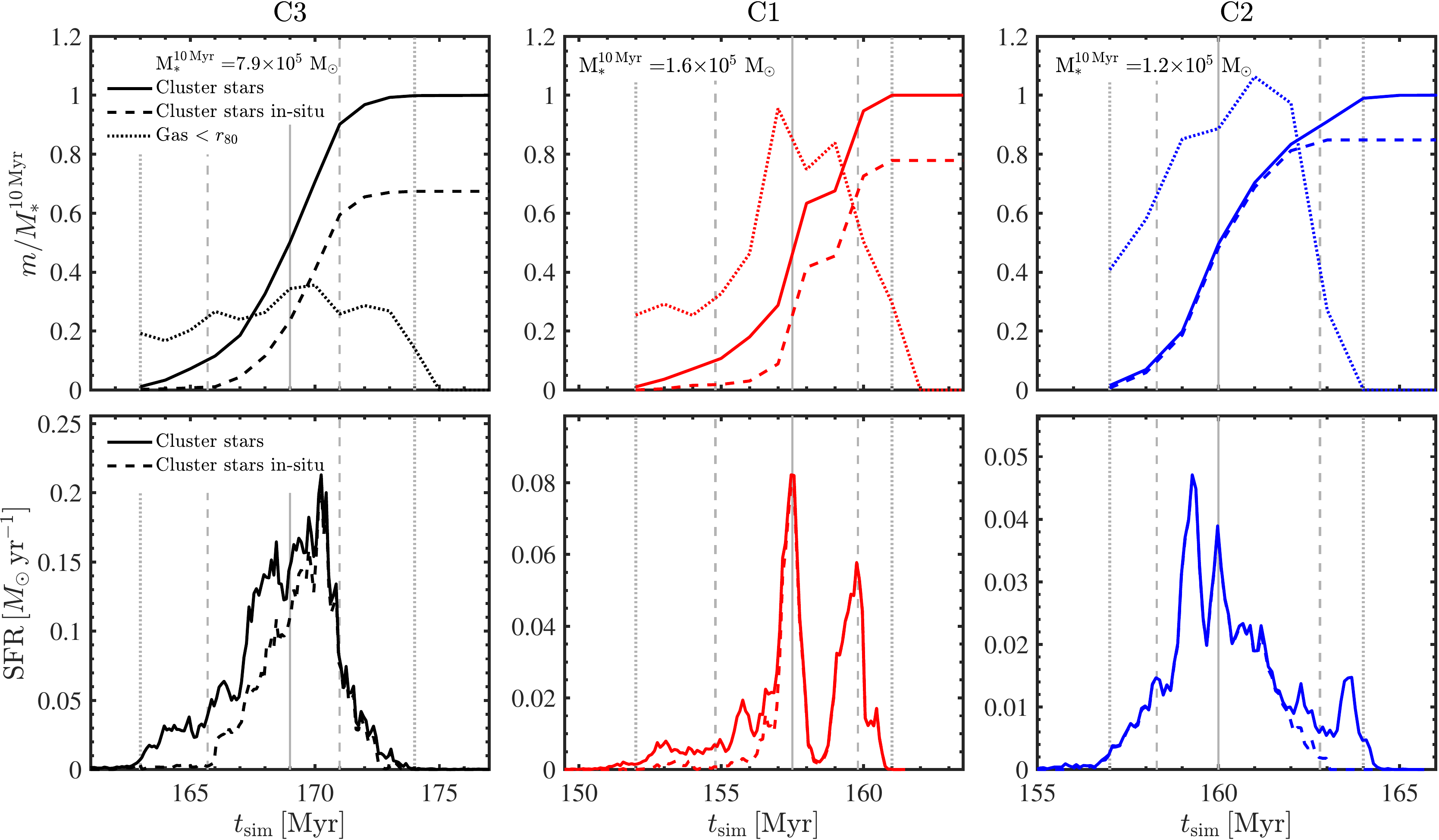}
\caption{The mass assembly histories (top) and star formation rates (bottom) of the three most massive clusters. The masses in the top row have been scaled by the bound mass at a cluster age of 10 Myr, and the stars have been separated into all cluster stars (solid lines) and cluster stars which formed within $r_{80}$ (dashed lines) of the local center of mass. The gas mass within $r_{80}$ is shown in the top row as a dotted line. All clusters are dominated by in-situ star formation with some contribution of smaller accreted star clusters. The SFRs have been calculated from the stellar ages of particles bound to each cluster at a mean stellar age of 10 Myr. The vertical lines show the epochs at which $50\%$ (solid) and $10\%$--$90\%$ (dashed) of the stellar mass has formed, and the time interval of cluster star formation in which we perform most of our analysis (dotted vertical lines). Note that the dashed and solid lines describing all cluster stars and in-situ cluster stars overlap for some of the time.  \label{fig:cumu_mass_sfr}}
\end{figure*}

\begin{figure}
\includegraphics[width=\columnwidth]{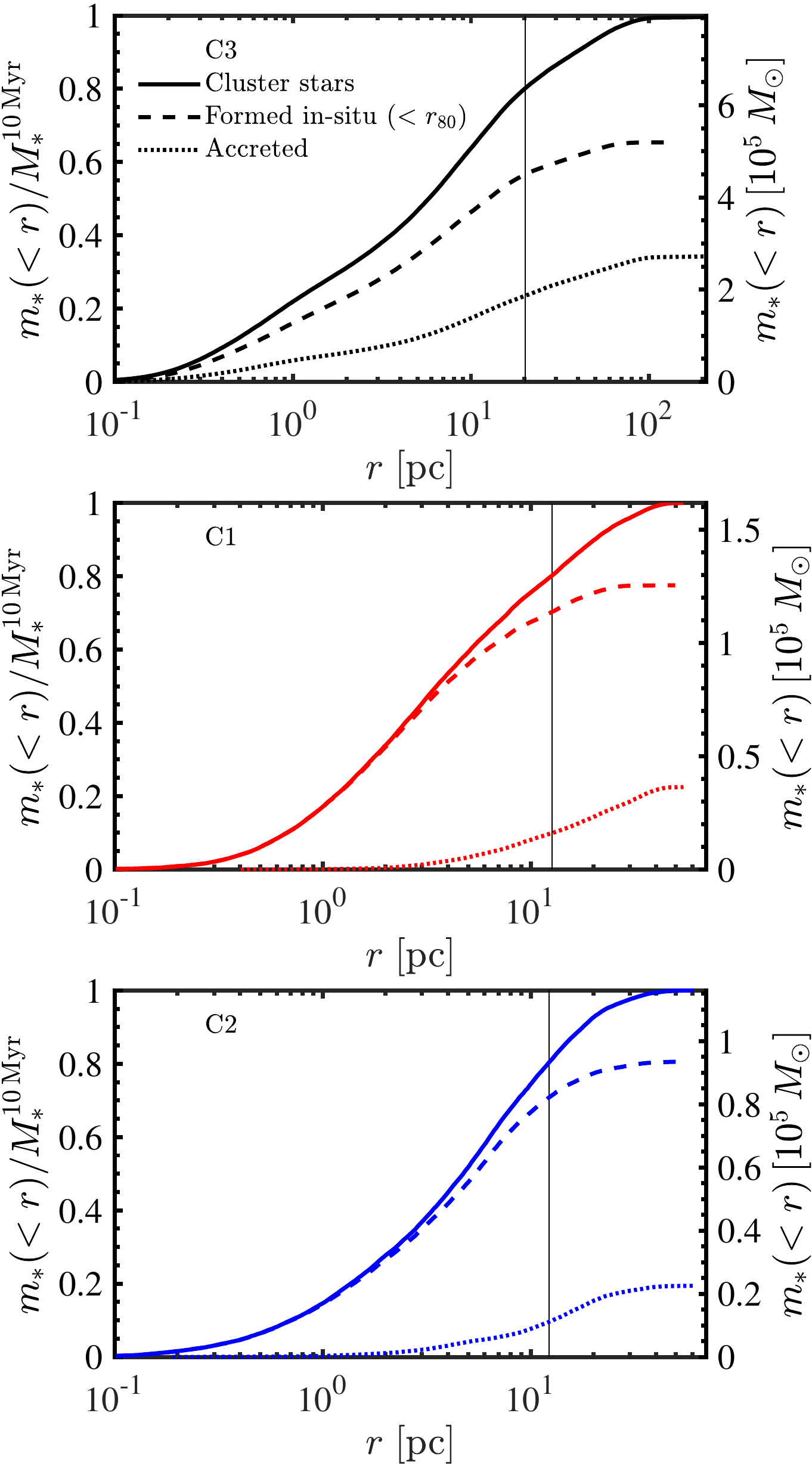}
\caption{The cumulative radial mass profiles of the three most massive clusters at a cluster age of 10 Myr. The solid line shows all bound stellar mass, the dashed line shows the mass formed within $r_{80}$, and the dotted line shows the accreted mass. The vertical lines show the value of $r_{80}$. Note the different extents of the radial coordinate on the x-axis.\label{fig:insitu_masses}}
\end{figure}

\section{Evolution of the most massive clusters}\label{sec:GCs}
\subsection{Assembly}\label{sec:assembly}

As shown in L20, a population of young star clusters forms and evolves with an observationally consistent power-law cluster mass function during the interaction of the two dwarf galaxies. In Fig. \ref{3_clusters} we show the main star formation region depicting the three most massive clusters forming in chronological order at around the time of the most intense starburst. Here the expulsion of the super shell by the first forming massive cluster, discussed in L19, is clearly visible. The gaseous environment evolves dramatically as the starburst kicks in. The most extreme regions in the disturbed gaseous disks in terms of thermal pressure, gas surface density and star formation surface density are the formation sites of the most massive star clusters. The numbering (C1--C3 in chronological order) indicated in Fig. \ref{3_clusters} is used to refer to the three massive clusters throughout this paper.

The clusters forming in Fig. \ref{3_clusters} are built from gas that resides within a couple hundred parsecs from the respective center of mass of each cluster formation region. Fig. \ref{fig:init_gas_radii} shows the distribution of radii of the gas that will end up constituting the bound stellar mass in each of the three clusters by the time they have reached a mean stellar age of 10 Myr. The epoch shown in Fig. \ref{fig:init_gas_radii} corresponds to the time when the star formation for each cluster begins in earnest, defined as the time of the last snapshot in which only a couple per cent of the cluster mass has formed. These times correspond to 163 Myr, 152 Myr and 157 Myr for C3, C1 and C2, respectively. Fig. \ref{fig:init_gas_radii} also shows the cumulative radial distribution of the initial gas mass. The initial gas distributions peak at a few tens of pc with median values between 20 and 50 pc, and 90\% of the gas mass originates from regions approximately 80 pc, 100 pc and 50 pc in size for C3, C1 and C2, respectively. The diminishing tail of the initial radii reaches out to few hundred parsecs. 

The initial gas cloud sizes in the rest of the young cluster population follow in general a decreasing trend with decreasing mass, with the majority of the lower mass clusters originating from gaseous regions less than 40 pc across. The results here are in line with the general picture of star cluster formation, where the most massive clusters form in mergers of smaller clusters rather than in a single collapse of a gaseous region as is the case with low mass star clusters \citep{2018NatAs...2..725H, 2020arXiv200600004C}.

\subsection{Formation}\label{sec:formation}

We investigate the build-up of the final bound stellar mass in the three clusters in Fig. \ref{fig:cumu_mass_sfr}. We follow the local gas content and the in-situ star formation using a fixed spherical region for each three clusters with an in-situ radius of $r_{80}$. The in-situ radius is defined as including $80\%$ of the stellar mass in each cluster once they reach a mean stellar age of 10 Myr. The in-situ radii ($r_{80}$) for the three clusters equal $20.2$ pc, $12.7$ pc, and $12.3$ pc for C3, C1 and C2, respectively, while the respective half-mass radii for the clusters are $5.9$ pc, $3.6$ pc and $4.8$ pc. The top row in Fig. \ref{fig:cumu_mass_sfr} shows the total (cumulative) mass of the cluster stars compared to the mass in the cluster stars formed in-situ, i.e. within $r_{80}$. The top row also shows the gas mass fraction within $r_{80}$ during the cluster formation process, which demonstrates that the cluster formation region is fairly gas-rich at least up to the time when half of the total cluster mass has formed.

The bottom row of Fig. \ref{fig:cumu_mass_sfr} shows the star formation rate produced by all stars which end up bound in the three clusters at an age of 10 Myr compared to the stars in the clusters which formed in-situ. We show here only the stars which will end up bound in the clusters, as this is the stellar archaeological variable which can be extracted from real star clusters even at old ages. The total time-dependent SFR is therefore defined as the stellar mass in age-bins of 0.1 Myr divided by the bin-width, extracted from the clusters individually when they reach an age of 10 Myr. The SFRs span from 0.01 to a couple of 0.1 $M_\odot\,\mathrm{yr}^{-1}$ and the star formation extends over a time span of the order of $10$ Myr. The prolonged star formation is partly fuelled by the gas inflow rate in these regions which are of the order of \mbox{$0.05$--$0.1\,M_\odot\,\mathrm{yr}^{-1}=0.5$--$1\times 10^5\,M_\odot\,\mathrm{Myr}^{-1}$}, which results in star formation efficiencies within the in-situ regions of the order of $25\%$-$50\%$. Similar levels of star formation and accretion rates have been found in molecular cloud scale simulations of star cluster formation \citep{2018NatAs...2..725H,2019MNRAS.487..364L,2020arXiv200600004C}.

The cumulative radial distribution of stellar mass is compared to the in-situ formed mass in Fig. \ref{fig:insitu_masses}. The clusters are dominated by the in-situ stars, and 60\%-70\% of the total mass in the clusters which originally formed in-situ still resides within $r_{80}$ when the clusters reach an age of 
10 Myr.

\subsection{Angular momentum}\label{section:angular_vector}

\begin{figure*}
\includegraphics[width=\textwidth]{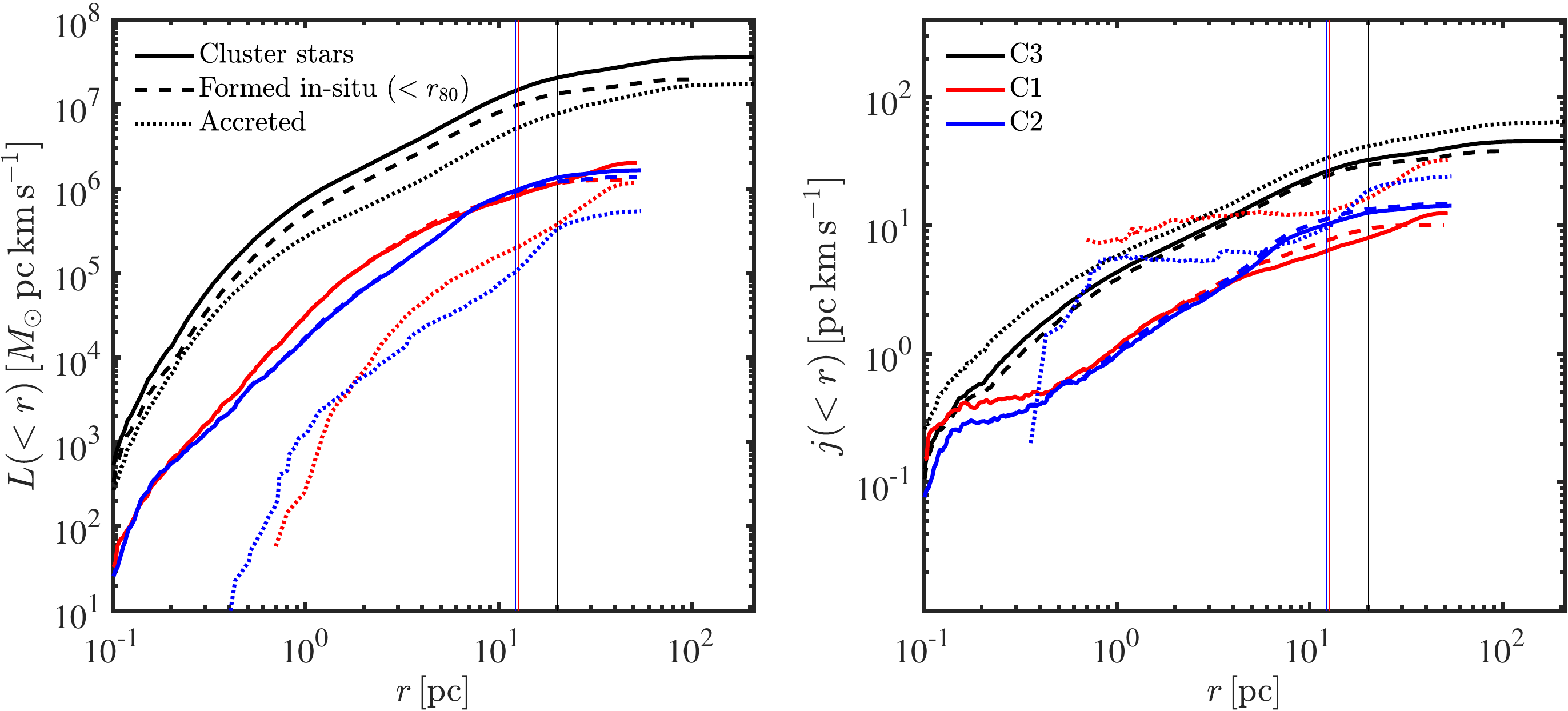}
\caption{The cumulative total (left) and the specific (right) angular momentum profiles of the three most massive clusters at an age of 10 Myr. The solid lines always show the total stellar mass, the dashed lines show the stellar mass formed in-situ ($<r_{80}$), and the dotted lines show the accreted stellar mass. The vertical lines show the value of $r_{80}$. The total angular momentum is dominated by in-situ stars, whereas the accreted stars have a higher specific angular momentum. \label{fig:cumulative_L}}
\end{figure*}

The cumulative radial distribution of angular momentum in the three most massive clusters C1-C3 are shown in Fig. \ref{fig:cumulative_L} at 10 Myr after their formation. The panels show the angular momentum and the specific angular momentum, with the data being separated into stars which formed in-situ and stars which have been accreted from outside of $r_{80}$. For the total angular momentum, the in-situ stars dominate the profile as they also dominate the mass according to Fig. \ref{fig:insitu_masses}. The specific angular momentum on the other hand is dominated by the accreted stars, as they come from larger radii and arrive originally with higher values of angular momentum.

\begin{figure*}
\includegraphics[width=\textwidth]{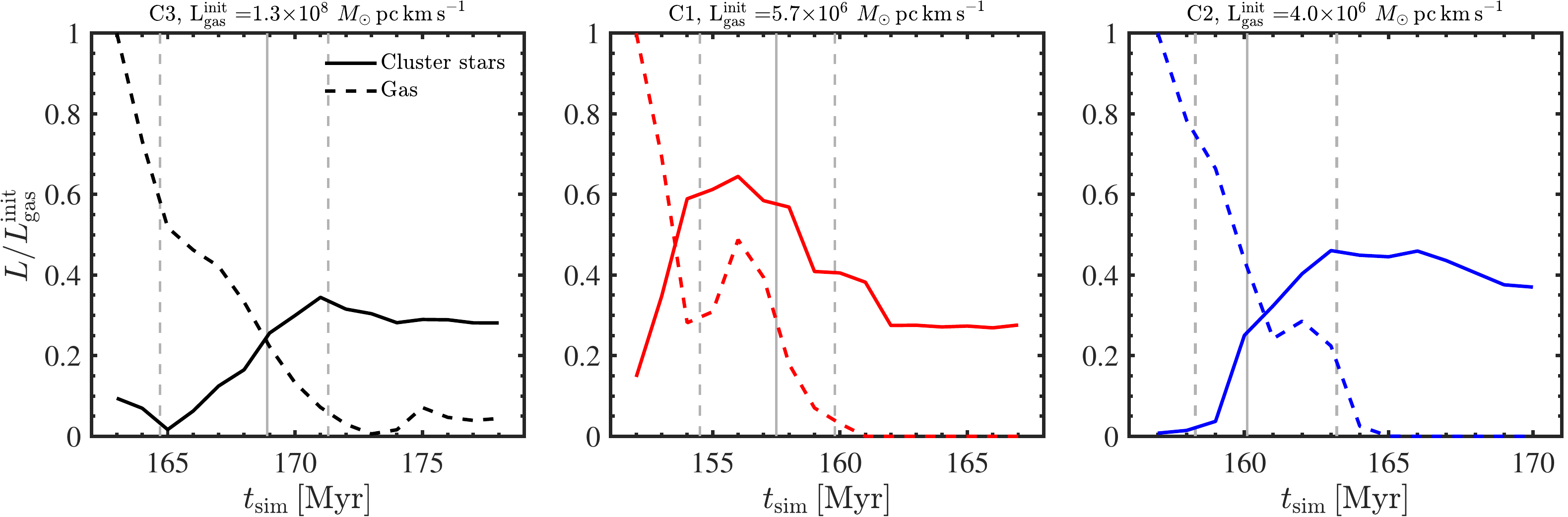}
\caption{The total angular momentum in gas and star particles which build up each of the three massive clusters in 1 Myr steps during their formation periods. The curves have been normalized to the initial value in gas in each cluster. The vertical lines indicate the epochs at which $10\%$, $50\%$ and $90\%$ of the final stellar mass has formed. This means that most of the gas has been consumed after the $90\%$ indicator and explains why the gas angular momentum profile rapidly declines as a function of time. 
\label{fig:tot_ang_mom}}
\end{figure*}

The simulation allows us to trace back the progenitor gas from which the stars of each individual cluster have formed. Here we again concentrate on particles which are bound to each cluster at the time when they reach an age of 10 Myr. In Fig. \ref{fig:tot_ang_mom} we show the total angular momentum, $L$, in the cluster stars as well as in the gas particles which will eventually turn into the cluster stars. The values have been scaled by the initial angular momentum of all progenitor gas particles in each cluster separately and calculated at the start of the star formation in each cluster (see e.g. Fig. \ref{fig:cumu_mass_sfr}). The angular momentum is shown during the cluster formation period (as indicated in Fig. \ref{fig:cumu_mass_sfr}) and up to the time when each cluster reaches an age of 10 Myr. The two most massive clusters already have a couple of per cent of their stars formed once we start the analysis, which have by definition formed ex-situ, and result thus in non-negligible angular momentum at the start of the analysis. All three clusters have roughly $30$\%--$40\%$ of their original gaseous angular momentum by the time they reach an age of 10 Myr.

Similar experiments using high resolution cloud-scale hydrodynamical simulations at a few orders of magnitude higher spatial resolution have been presented in  e.g.  \citet{2016A&A...591A..30L} and \citet{2017MNRAS.467.3255M}.  For example in \citet{2017MNRAS.467.3255M} all clusters except those with stellar masses below $100\, M_\odot$ showed signatures of ordered rotation, even though their initial conditions were non-rotating. \citet{2016A&A...591A..31L} in turn showed how the properties of the forming protoclusters can be derived from the initial star-forming cloud. The high level of turbulence and strong tidal torques commonly found in the high-pressure environments where star clusters tend to form, such as starbursts and mergers studied here, makes it difficult to form a non-rotating massive star clusters due to the conservation of angular momentum.

Finally, we briefly investigate the direction of the angular momentum vector in the three clusters with respect to the center of mass of each cluster during their initial formation and at an age of 100 Myr. The purpose is to test whether the clusters end up rotating in the same direction as the progenitor gas or if the direction of the cluster rotation is instead set by the global galactic encounter orbit. The most massive cluster (C3) forms out of gas which rotates more or less in the plane of global rotation, and the cluster ends up rotating in a similar direction. The two less massive clusters (C1 and C2), on the other hand, form from gas which ends up rotating in the opposite direction. The direction of the momentum vector in the C1 progenitor gas is torqued towards the opposite direction during the star formation process while the C2 progenitor gas already rotates in the opposite direction with respect to the global merger orbit when the cluster formation begins. 

Studying the rotational axis of the nine most massive clusters, four rotate in the same (but misaligned) direction as the global angular momentum and the rest rotate in opposite (but not antiparallel) direction, whereas the lower mass clusters span randomly the entire range of rotational directions. Combining the directional information with the total fraction of angular momentum shown in Fig. \ref{fig:tot_ang_mom}, the star clusters clearly inherit the local angular momentum properties of the turbulent progenitor gas as already reported in isolated cloud-scale simulations of cluster formation (e.g. \citealt{2017MNRAS.467.3255M}). On the other hand, in Fig. \ref{fig:misalignment} we showed how the $\sim9$ most massive clusters are the least misaligned with respect to their shapes. The absolute amount of angular momentum is significant enough to result in clear aligned rotational signal only in the most massive clusters even though the lower mass clusters also have non-zero net angular momentum (as in Fig. \ref{fig:specific_j}).

\begin{figure*}
\includegraphics[width=0.95\textwidth]{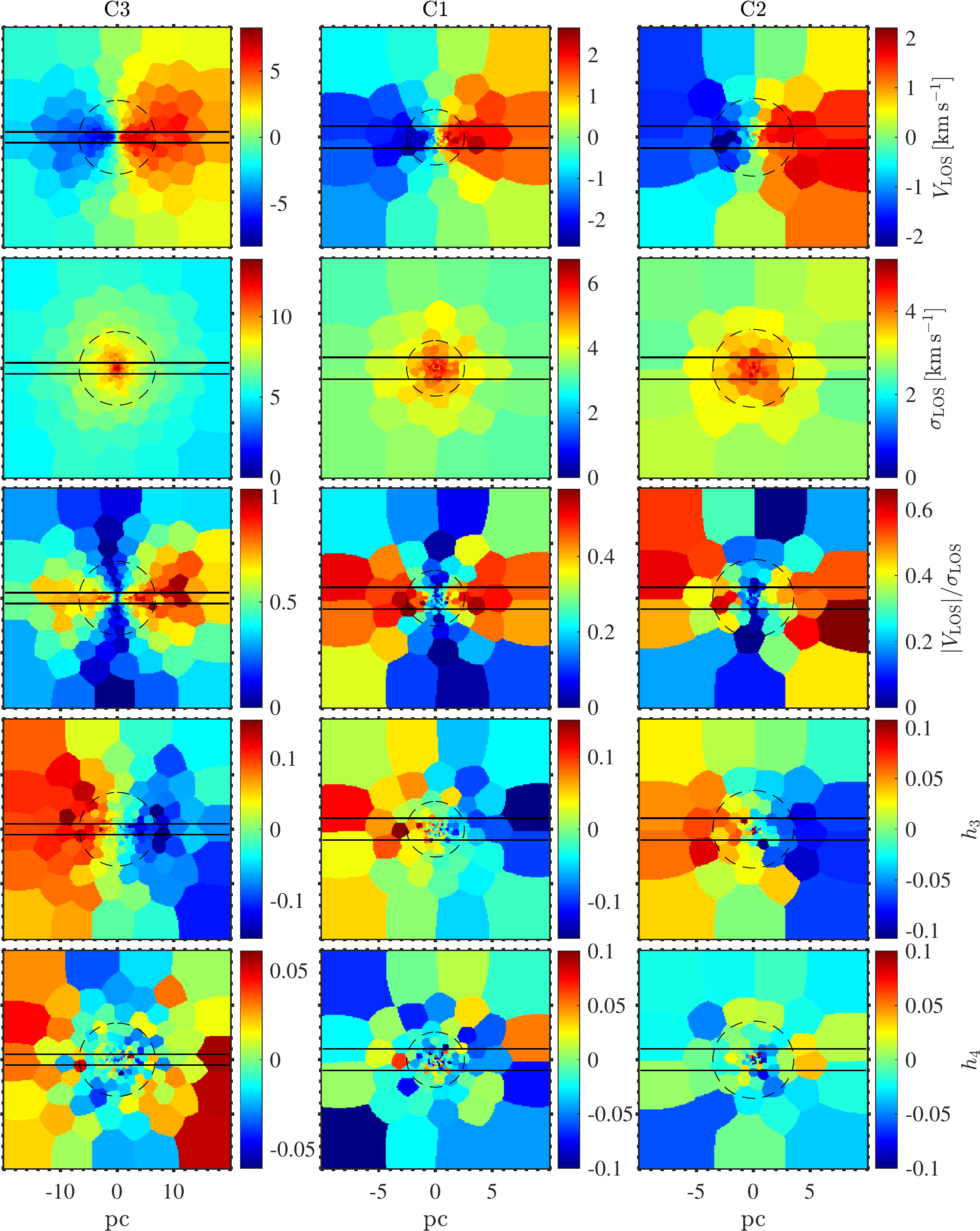}
    \caption{The line-of-sight mass weighted mean velocity, velocity dispersion and $V/\sigma$ perpendicular to the angular momentum vector (i.e. edge-on), as well as the higher order Gauss-Hermite parameters $h_3$ and $h_4$ in the three most massive clusters 100 Myr after their formation. The cluster mass decreases from left to right and the cluster identification numbers from Fig. \ref{3_clusters} are indicated in the top panels. The horizontal lines indicate the width of the slit used to calculate the radial profiles and the circles show the region within the half-mass radius. All Voronoi cells include at least 100 particles. The clusters show a clear anti-correlation between velocity and $h_3$ with tentatively negative central values of $h_4$. Note the different extents and color ranges in the different panels. \label{fig:velocities}}
\end{figure*}

\begin{figure}
\includegraphics[width=\columnwidth]{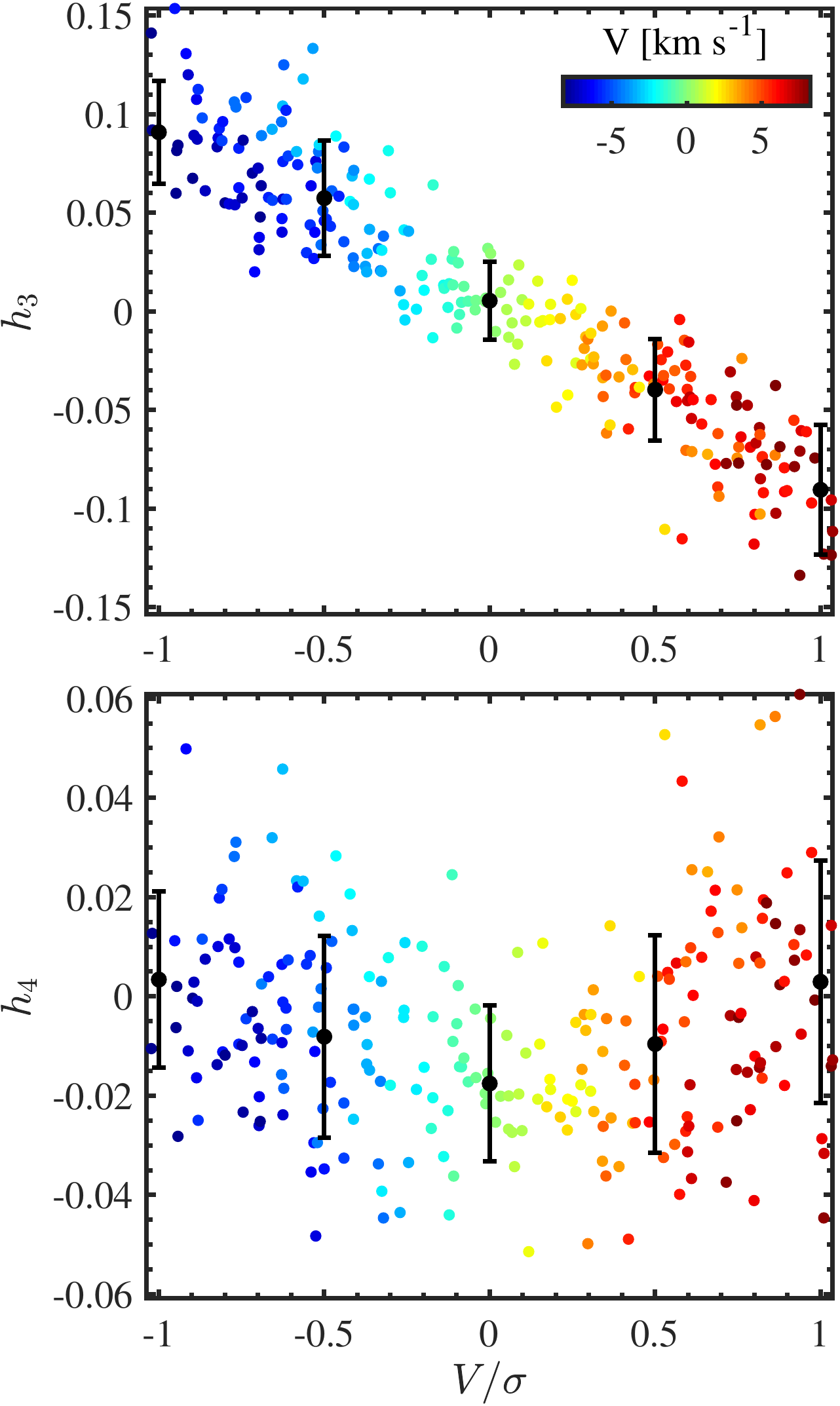}
\caption{Gauss-Hermite parameters $h_3$ (top) and $h_4$ (bottom) for each Voronoi spaxel (see Fig. \ref{fig:velocities}) from fitting the stellar line-of-sight velocity distribution vs. $V/\sigma$ in the edge-on projection for the most massive cluster C3 at an age of 100 Myr. The errorbars show the mean and standard deviation of the data points in five bins. The clear $h_3$ - $V/\sigma$ anti-correlation and negative $h_4$ values at low $V/\sigma$ are a dynamic signature for oblate rotation \citep[see e.g.][]{1993MNRAS.265..213G}. \label{fig:gauss}}
\end{figure}

\begin{figure*}
\includegraphics[width=\textwidth]{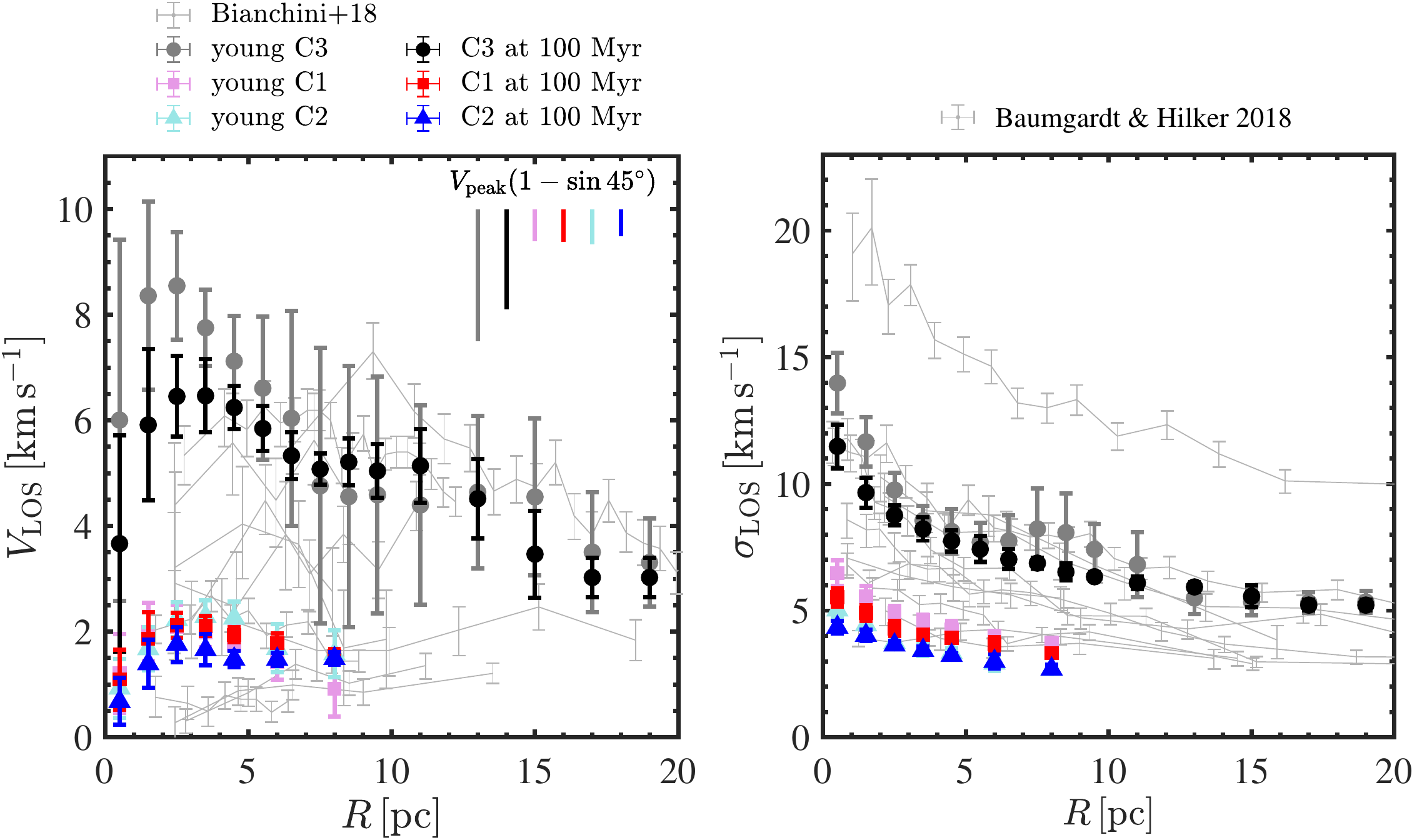}
\caption{Radial profiles of the line-of-sight velocity (left) and the velocity dispersion (right) perpendicular to the angular momentum vector in the three most massive clusters immediately after (fainter symbols) and 100 Myr after the starburst (darker symbols). The radial profiles have been calculated over a $2$ pc slit along the plane of rotation (see fig. \ref{fig:velocities}). The bars in the top right corner of the left hand panel show the reduction in the value of the peak velocity of each simulated profile if it were observed at a $45^\circ$ inclination. The errorbars show the bootstrapped standard deviations. The observed profiles for 11 Milky Way GCs with masses between $1.88\times 10^5$--$3.55\times 10^6\,M_\odot$ from \citet{2018MNRAS.481.2125B} (left) and \citet{2018MNRAS.478.1520B} (right) are shown underneath. The uppermost observed velocity dispersion profile represents the $\omega$ Cen. \label{fig:velocity_prof}}
\end{figure*}

\begin{figure}
\includegraphics[width=\columnwidth]{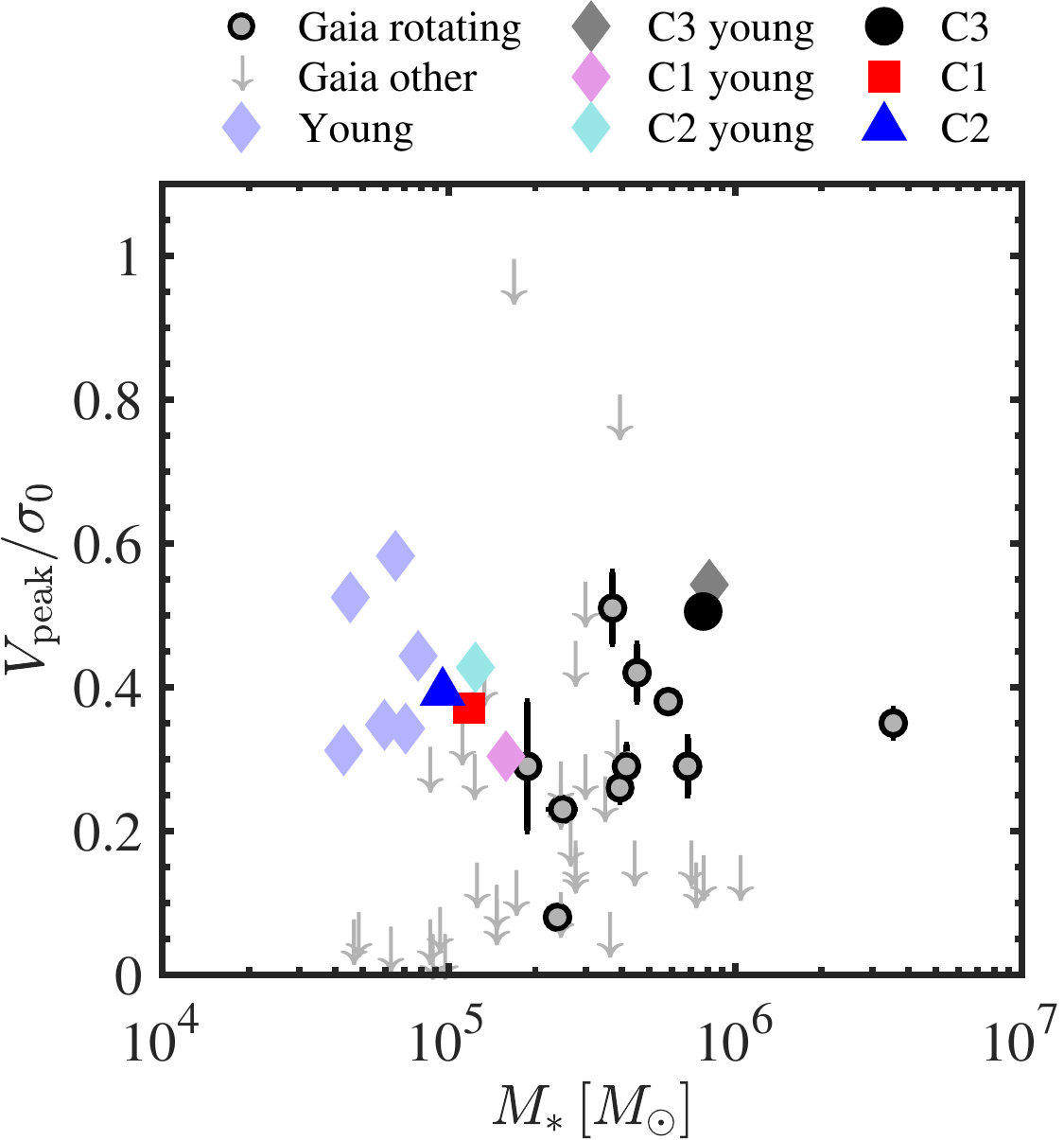}
\caption{The peak velocity within $2r_\mathrm{h}$ per central velocity dispersion, $V_\mathrm{peak}/\sigma_0$. The radial extent of $V_\mathrm{peak}$ is limited to within $2r_\mathrm{h}$ following the typical maximum extent reached in observations. The three most massive clusters have been highlighted with the same colors as used in Fig. \ref{fig:velocities} right after the starburst (diamonds) and at a mean stellar age of 100 Myr, and the light blue diamond symbols show the other fairly massive rotating clusters right after the starburst. The Gaia DR2 data points for old Milky Way GCs are from Table 1 of \citet{2018MNRAS.481.2125B}. The data points of the clusters shown in Fig. \ref{fig:velocities} are highlighted with black and the clusters with only upper limits reported in \citet{2018MNRAS.481.2125B} are shown as downward arrows. Note that all the observed Milky Way GCs have ages in excess of 10 Gyr. \label{fig:vpersigma}}
\end{figure}

\section{The line-of-sight velocity distribution}\label{section:los_maps}

\subsection{Projected velocity maps}

In Fig. \ref{fig:velocities} we show the projected edge-on line-of-sight velocities ($V_\mathrm{LOS}$), velocity dispersions ($\sigma_\mathrm{LOS}$) and $V_\mathrm{LOS}/\sigma_\mathrm{LOS}$ for the three most massive clusters out to a few half-mass radii. In addition we also fit Gauss-Hermite functions and give the higher order coefficients $h_3$ and $h_4$ \citep[see e.g.][]{1993MNRAS.265..213G} in the two bottom rows of Fig. \ref{fig:velocities}. The fitting procedure is described in detail in \citet{2014MNRAS.444.3357N} only that here we directly fit on the velocities of the individual stars (see Sec. \ref{sec:kinematics}). We study the kinematics after several half-mass relaxation times  for better comparison to local, much older,  globular clusters. 

The anti-correlation between $V_\mathrm{LOS}$ and $h_3$ for the three clusters is already apparent form  Fig. \ref{fig:velocities} and highlighted for the most massive cluster in Fig. \ref{fig:gauss}. It indicates a steep leading wing in the velocity distribution and can be reproduced in models of isotropic (see Fig. \ref{fig:beta}) rotators \citep{1994MNRAS.268.1019D} or more complex configurations such as rotating disks embedded in non-rotating spherical stellar distributions \citep[see e.g.][]{2001ApJ...555L..91N}. We also see in Fig. \ref{fig:velocities} and Fig. \ref{fig:gauss} indication of slightly negative central values of $h_4$ which describe flatter than Gaussian line-of-sight velocity distributions. In galaxy formation models it has been shown that the anti-correlation between $h_3$ and $V/\sigma$ is a clear kinematic signature of gas dissipation and in-situ star formation \citep[see e.g.][]{2006MNRAS.372..839N,2014MNRAS.444.3357N,2014MNRAS.445.1065R}. This is very similar to the clusters shown here (Section \ref{sec:formation}) and it remains to be seen whether observed massive star clusters or globular clusters show similar signatures. 

Our clusters are young compared to old globulars $(\sim 10\, \rm Gyr)$ and it is expected that the simulated clusters would further lose angular momentum as they evolve in a galactic tidal field. 
$N$-body simulations (e.g. \citealt{2013MNRAS.430.2960H} and \citealt{2017MNRAS.469..683T}) have  studied the loss of angular momentum of tidally limited star clusters and conclude that the angular momentum needs several relaxation time scales to decrease significantly. They also found that some angular momentum remains even after tens of relaxation times, resulting in a non-negligible $V/\sigma$ value. We could expect at least the most massive simulated clusters to retain some amount of their angular momentum even after Gyrs of evolution, in a similar fashion to what is seen in present-day massive GCs, which show measurable levels of rotation even at an age in excess of 10 Gyr. 

The loss of angular momentum is attributed to both stars escaping due to the surrounding tidal field and to stellar mass loss through feedback. All of our three clusters form a central core (see Fig. 4 in L19) and the clusters lose $10\%$--$30\%$ of their total bound mass during their first 100 Myr of evolution. When we look at the radii enclosing a given mass ($10\%$, $50\%$ and $80\%$ of the final cluster mass) all the radii increase at least by $\sim30\%$ and the radii which enclose $80\%$ of the final cluster mass at least double in size during the 100 Myr of evolution. On the other hand, for example the radii which enclose $50\%$ of stars \textit{by number} also increase by $10$--$20\%$. The number of stars within fixed radii, e.g. $r_{80}$ and $r_h$, also decrease up to $\sim 10\%$ as the clusters evolve (compared to up to $25\%$ when considering stellar mass), indicating also true escapers rather than simple mass loss through winds. The mass loss would then continue until the clusters evaporate, however properly resolving this process would require a more
accurate integration method than our current softened hydrodynamical simulation.

\subsection{Radial line-of-sight velocity profiles}

To quantify the rotational properties we calculate radial profiles for the LOS velocity and velocity dispersion maps from Fig. \ref{fig:velocities}. The profiles are shown in the left and right panels of Fig. \ref{fig:velocity_prof}, and are compared to the velocity and dispersion profiles of 11 old Milky Way GCs in the mass range between $1.88\times 10^5$--$3.55\times 10^6\,M_\odot$ from \citet{2018MNRAS.481.2125B} and \citet{2018MNRAS.478.1520B}. The observed profiles have been obtained by averaging the tesselated (inclined) tangential motion in a face-on manner. We follow this approach by observing our clusters from a direct edge-on point of view. The LOS velocity profiles were calculated in a $2\, \rm pc$ wide slit along the plane of rotation (indicated in Fig. \ref{fig:velocities}), similar to what was used in e.g. calculating the estimates for the de-inclined tangential velocities in \citet{2018MNRAS.481.2125B}. We perform the same analysis and calculated the velocity profiles also immediately after the starburst and confirm the slow down of the clusters under the influence of the galactic tidal field also seen in detailed direct $N$-body simulations.

All three clusters show clear ordered rotation at a rate of a few $\rm{km}\,\rm{s}^{-1}$ even after 100 Myr of evolution. The $V/\sigma$ values peak approximately at $1$--$2$ half-mass radii with values in the range $\sim0.65-1$ as seen in $N$-body simulations \citep{2014MNRAS.443L..79V,2017MNRAS.469..683T}. The rotational curves of the 100 Myr old most massive and two less massive clusters peak at $\sim 6.5 \, \mathrm{km}\, \mathrm{s}^{-1}$ and $2.1$--$1.8 \, \mathrm{km}\, \mathrm{s}^{-1}$, and have values at half-mass radius of $\sim 5.2 \, \mathrm{km}\, \mathrm{s}^{-1}$ and $2.1$--$1.6 \, \mathrm{km}\, \mathrm{s}^{-1}$, respectively. Galactic globular clusters with similar masses and significant rotation have similar peak rotation with values up to $\sim 6 \, \mathrm{km}\, \mathrm{s}^{-1}$ \citep{2018MNRAS.481.2125B}. The simulated clusters show however more centrally concentrated velocity profiles compared to the observed globulars where the angular momentum has had a much longer time for outwards redistribution. For strongly cored stellar systems, i.e. for flat constant density cores such as those observed in evolved GCs, the velocity dispersion should also tend to zero at small radii (see e.g. \citealt{1993MNRAS.265..250D}).

The simulated radial profiles in Fig. \ref{fig:velocity_prof} are shown in an edge-on projection, whereas observed velocity measurements can rarely be obtained strictly along or perpendicular to the rotational axis. \citet{2018MNRAS.481.2125B} discuss the effect of inclination on the clusters shown on the background of Fig. \ref{fig:velocity_prof}, reporting values from $75^\circ$ downwards. We approximate the effect of an inclined line-of-sight by showing in Fig. \ref{fig:velocity_prof} the decrease in the peak value of each modelled profile caused by a $45^\circ$ inclination. Such an inclined rotation would cause a $\sim2\,\mathrm{km\,s^{-1}}$ decrease in the peak value of the most massive cluster at an age of 100 Myr, while the two less massive clusters would be seen to rotate at a $\sim0.6\,\mathrm{km\,s^{-1}}$ slower rate. As a result, were we to adjust the observed profiles upwards to correct for inclination or to adjust the simulated profiles downwards to account for inclination, the match between the observed and simulated profiles would be even better.

Fig. \ref{fig:vpersigma} shows the peak LOS velocity per central velocity dispersion $V_\mathrm{peak}/\sigma_0$ in the nine massive clusters highlighted in Fig. \ref{fig:specific_j} obtained by producing similar radial LOSVD profiles for all the massive clusters as shown in Fig. \ref{fig:velocity_prof}. Here we show the nine clusters right after the starburst, and overplot the three massive GC candidates at an age of 100 Myr. The observed data on the background shows all the measurements from Table 1 of \citet{2018MNRAS.481.2125B}. The 11 Milky Way GCs shown in Fig. \ref{fig:velocity_prof} with clear rotation reported in the velocity measurements and $V_\mathrm{peak}/\sigma_0$ error estimates have been highlighted with black. The $V_\mathrm{peak}/\sigma_0$ parameter is of the order of $0.5$ for the most massive cluster and $0.3$--$0.6$ for the less massive clusters, compared to observations which span typically from $0.1$ to $0.5$ for the rotating clusters \citep{2018MNRAS.481.2125B}. As we saw in Fig. \ref{fig:velocity_prof}, the peak velocities are higher for our less evolved clusters, whereas the central velocity dispersion values are in better agreement with the observational data. The result is higher $V_\mathrm{peak}/\sigma_0$ values compared to the old MW GCs. The effect of inclination in the observed values might on the other hand increase the intrinsic values of the observed clusters. Note also how some of the observed clusters without significant rotation can also have significant upper limit values for $V/\sigma$. In view of the inherent uncertainties of the $V_\mathrm{peak}/\sigma_0$ diagnostic, and the tentative agreement of the radial profiles, the results obtained here are very encouraging for future studies of simulated star cluster kinematics.

\begin{figure*}
\includegraphics[width=\textwidth]{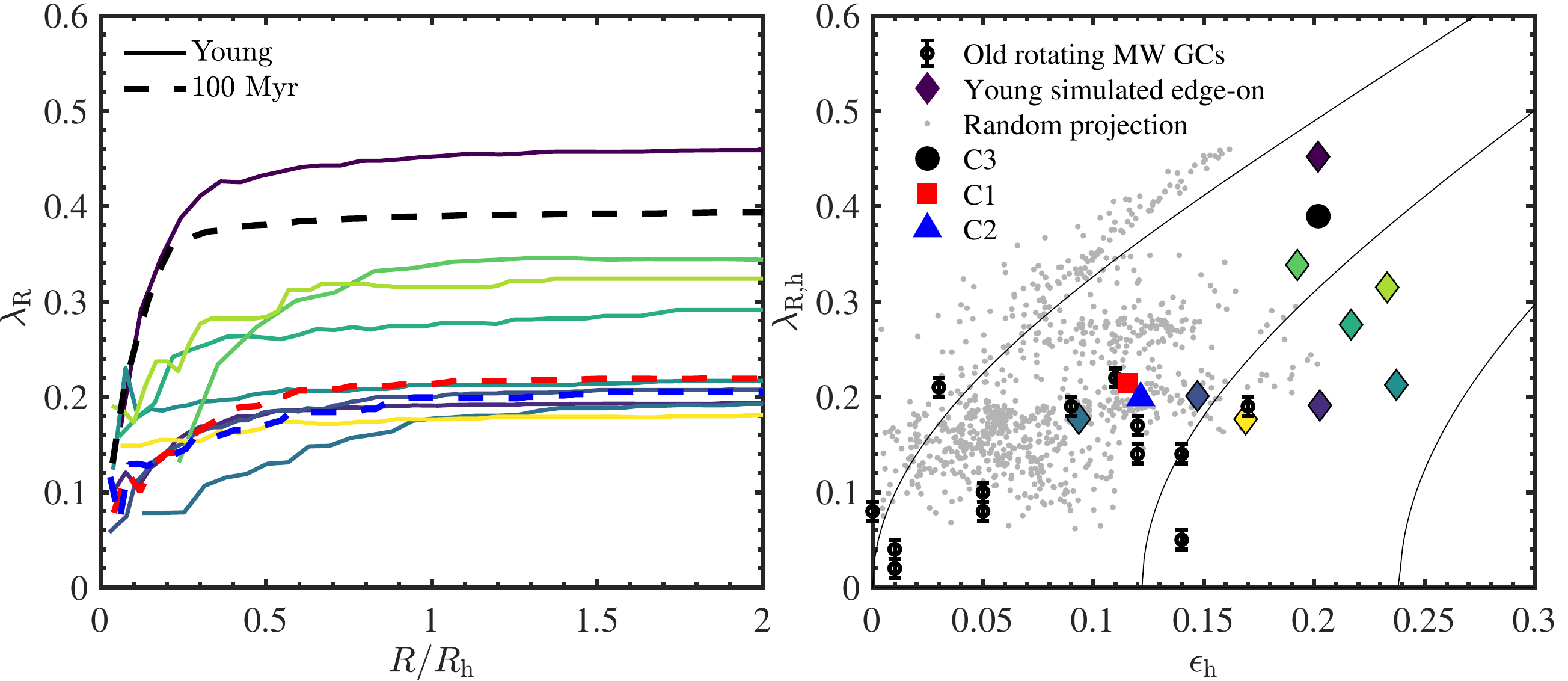}
\caption{Radial profiles of the dimensionless $\lambda_\mathrm{R}$ parameter (left) and the $\lambda_\mathrm{R}$ value as a function of intrinsic ellipticity at half-mass radius (right) in the nine analysed massive star clusters right after the starburst, as well as in the three most massive clusters once they reach 100 Myr in mean stellar age. The stellar masses  in the young clusters increase with the darkness of the lines (datapoints) from yellow through green to purple and the old clusters are depicted with the same symbols as in Fig. \ref{fig:vpersigma}. The small data points on the right show the projected values from 100 random lines-of-sight for each young cluster and the lines indicate the relation between $\epsilon$ and $\lambda_\mathrm{R}$ for an edge-on viewed oblate rotator with velocity anisotropy values of 0.0, 0.1 and 0.2 (as expressed in Fig. \ref{fig:dispersion}). The observed datapoints from \citet{2018MNRAS.473.5591K} and \citet{2010arXiv1012.3224H} show clusters with significant rotational signal identified in \citet{2018MNRAS.473.5591K}.
\label{fig:lambda_R}}
\end{figure*}

\subsection{Line-of-sight angular momentum}

The mass-weighted (or luminosity-weighted) angular momentum along the LOS can also be obtained from the projected velocity maps. A typical measure for this is the dimensionless angular momentum parameter $\lambda_{R}$ \citep{Emsellem2007, Emsellem2011}, defined as 
\begin{equation}\label{eq:lambda}
    \lambda_\mathrm{R}=\frac{\sum_{i=1}^N F_i R_i |V_{\mathrm{LOS},i}|}{\sum_{i=1}^N F_i R_i \sqrt{V_{\mathrm{LOS},i}^2+\sigma_{V,i}^2}}
\end{equation}
where $F_i$ and $R_i$ denote the mass (in observations the luminosity) and the radius of a given pixel $i$. The sums go over all pixels within a given radius. $\lambda_\mathrm{R}$ is by nature cumulative, and it has been shown to robustly characterise the global rotation in early-type galaxies \citep{2009MNRAS.397.1202J}, better than for example the $V/\sigma$ diagnostic. The dimensionless $\lambda_\mathrm{R}$ parameter is typically used to characterise early-type galaxies into categories of fast and slow rotation based on the 
threshold value of $\lambda_\mathrm{R}=0.1$, typically measured at the half-light radius. Objects above this threshold often show clear rotational signal while for objects below this value the velocity distribution is mainly dominated by dispersion.

Recently with detailed IFU spectroscopy $\lambda_\mathrm{R}$ has also been introduced in the context of smaller mass objects such as star clusters.
We produce the radial $\lambda_\mathrm{R}$ profiles for all nine massive clusters highlighted in Fig. \ref{fig:specific_j} and show the results in Fig. \ref{fig:lambda_R}. The left panel of Fig. \ref{fig:lambda_R} shows the $\lambda_\mathrm{R}$ profiles and the right hand panel quantifies the $\lambda_\mathrm{R}$ values at the half-mass radius with respect to the intrinsic ellipticity. The $\lambda_\mathrm{R}$-values at half-light radii for globular clusters with rotation obtained from \citet{2018MNRAS.473.5591K} (NGC 104, NGC 1851, NGC 2808, NGC 5139, NGC 5904, NGC 6093, NGC 6266, NGC 6293, NGC 6388, NGC 6541, NGC 6656, NGC 7078 and NGC 7089) are shown in the background of the right hand panel of Fig. \ref{fig:lambda_R}, with ellipticities at the half-light radius obtained from \citet{2010arXiv1012.3224H}. The curves in the right hand panel of Fig. \ref{fig:lambda_R} show the $\lambda_\mathrm{R}$-$\epsilon_\mathrm{h}$ relation of an edge-on oblate rotator with isotropic velocity distribution (left) and with velocity anisotropies of 0.1 and 0.2 using the formalism in Fig. \ref{fig:dispersion} and equations derived in \citet{2007MNRAS.379..418C} and \citet{2007MNRAS.379..401E}.

The simulated clusters reach mostly their maximum $\lambda_\mathrm{R}$ values within $1.5\times r_\mathrm{h}$. The (intrinsic) values at the half-mass radius span $0.18$--$0.45$, indicating that they all exhibit at least some level of rotation. On the other hand the observed datapoints also show very low values of  $\lambda_\mathrm{R,h}$ even though \citet{2018MNRAS.473.5591K} report significant rotation.

We investigate the effect of inclination by producing the LOSVD maps and $\lambda_\mathrm{R}$ profiles along 100 random lines-of-sight. The resulting 900 $\lambda_\mathrm{R,h}$ and projected $\epsilon_\mathrm{h}$ pairs are shown in the background of Fig. \ref{fig:lambda_R}. The random projections show how the projected ellipticity is always smaller than the intrinsic ellipticity, as for example can be analytically expressed in the case of an oblate ellipsoid
\begin{equation}
    \epsilon_\mathrm{intr}=1-\sqrt{1+\epsilon_\mathrm{app}(\epsilon_\mathrm{app}-2)/\sin^2(\iota)}
\end{equation}
where $\epsilon_\mathrm{intr}$ and $\epsilon_\mathrm{app}$ are the intrinsic the apparent ellipticities and $\iota$ is the inclination \citep{2007MNRAS.379..418C}. The same goes for the dimensionless angular momentum, as the separately highlighted edge-on projections always show the most extreme angular momentum profiles attainable from the data. The most massive cluster and its random projections are consistent with the line for isotropic oblate rotator, as also seen in the anisotropy measurements in Figures \ref{fig:dispersion} and \ref{fig:beta}.  

Finally, we briefly tested the effect of resolution by doubling and halving the number of Voronoi cells in the production of the LOSVD maps. The typical variation in $\lambda_\mathrm{R,h}$ introduced by the spaxel resolution is of the order of or less than $\pm 0.03$.

\begin{table*}
\begin{center}
\begin{tabular}{| c c c c |}
    \hline
     & C3 & C1 & C2\\ 
     \hline
    $t_\mathrm{sim}$ at SF start and SF shut off [Myr] & [163, 174] & [152, 161] & [157, 164] \\
    mean gas inflow\footnote{Into $r_{80}$} [$M_\odot$ yr$^{-1}$] & 0.138 & 0.047 & 0.049 \\
    \hline
    Measured at cluster age $\sim 10$ Myr & & &\\
    $M_*$ [$10^5\, M_\odot$] & $7.9$ & $1.60$ & $1.23$ \\
    $r_{h}$ [pc] & $5.9$ & $3.6$ & $4.8$ \\
    $r_{80}$ [pc] & $20.2$ & $12.7$ & $12.3$ \\
    $f_{*\mathrm{,\,in-situ}}$\footnote{fraction of $M_{*}$ formed within $r_{80}$} & $0.65$ & $0.78$ & $0.81$ \\
    $f_{*\mathrm{,\,in-situ}}(<r_{80})$\footnote{fraction of $M_{*}$ formed within $r_{80}$ still within $r_{80}$}  & $0.57$ & $0.70$ & $0.71$ \\
    \hline 
    Measured 100 Myr after the starburst: & & & \\
    $M_*$ [$10^5\, M_\odot$] & $7.69$ & $1.18$ & $0.95$ \\
    $R_{h}$ [pc] & $6.2$ & $2.5$ & $3.6$ \\
    $V_{\mathrm{LOS},\,h}$\footnote{Within $r_\mathrm{h}\pm0.5$ pc}\footnote{Standard deviations based on the Voronoi-binned data in Fig. \ref{fig:velocities}} [km s$^{-1}$] & $5.2\pm 0.4$ & $2.1\pm 0.2$  & $1.6\pm 0.3$ \\
    $\sigma_\mathrm{h}$ [km s$^{-1}$] & $7.0\pm 0.4$ & $4.3\pm 0.3$ & $3.4\pm 0.2$ \\
    $\sigma_0$\footnote{Within $r<0.5$ pc} [km s$^{-1}$] & $12.8\pm 0.4$ & $5.7\pm 0.5$ & $4.5\pm 0.3$ \\
    $V_\mathrm{peak}/\sigma_0$\footnote{$V_\mathrm{peak}$ based on the $V_\mathrm{LOS}$ profile in Fig. \ref{fig:velocity_prof}} & $0.51$ & $0.37$ & $0.39$ \\
    $\lambda_\mathrm{R,h}$& $0.39$ & $0.21$ & $0.20$ \\
    \hline
    \end{tabular}
    
\caption{The properties of the three most massive star clusters $\sim 10$ Myr and $\sim100$ Myr after their formation.}
\label{table:t1}
\end{center}
\end{table*}

\section{Conclusions}

We have analysed the intrinsic shape and angular momentum properties of the entire young star cluster population in the mass range $M_{\mathrm{cl}} = 2 \times 10^2 M_\odot - 9 \times 10^5 M_\odot$ formed in a high-resolution simulation of a starburst triggered by a low-metallicity gas-rich dwarf galaxy merger. In addition to many low mass clusters the simulations also contains a few young massive star clusters which are plausible progenitors of dense and massive globular cluster systems. The massive star clusters form from gas which is mostly located within $\sim 100 \ \rm pc$ of the local center of mass, but a few per cent of the cluster mass may originate from distances in excess of 200 pc. Smaller mass clusters form from smaller regions, typically less than a few tens of parsecs across.

The young cluster population forms with a mass-dependent specific angular momentum distribution in which the specific angular momenta $j_\mathrm{h}$ measured within the half-mass radius $r_\mathrm{h}$ increase with cluster mass with a sub-linear relationship of $j_\mathrm{h}\propto M^{0.62}$ consistent with simple expectations from molecular cloud collapse. Even the least massive clusters have a non-zero $j_\mathrm{h}$ with slight tapering at $0.01$--$0.1$ pc km s$^{-1}$, albeit with a larger scatter. The most massive clusters, in which we see a clear rotational signal, have the least elongated ($\epsilon_\mathrm{h}\sim0.2$) but most oblate shapes. Their shapes (using here the shortest principal axis within $r_\mathrm{h}$) are typically aligned within $<20$ degrees of their angular momentum vector. The specific angular momenta in the most massive clusters are also dominated by accreted stars at all radii where accreted stars are found. 

The lower mass clusters, on the other hand, do not show a clear correlation between their shapes and angular momentum vectors. In general, they have lower values of specific angular momentum, and yet they seem more elongated in their shapes. Contrary to massive stellar objects where flattening is often explained with rotation, the flattened shapes of the less massive star clusters cannot be in our study attributed to rotation. The elliptic shapes are more probably a result of filamentary collapse, or external effects such as the general tidal field and torques from other nearby matter.

We pay special attention to the three most massive clusters ($\gtrsim 10^5\,M_\odot$), which evolve towards properties very similar to observed present-day globular clusters in the Local Group. We collect some of the main parameters obtained in this study for the three globular cluster candidates in Table \ref{table:t1}. The three clusters form their bound mass during a time span of $\sim 10$ Myr, which results in age spreads of the order or less than $\sim5$ Myr. The SFRs in the cluster formation regions reach peak values of up to $0.2\,M_\odot\,\mathrm{s}^{-1}$ with time-averaged SFR of the order from $0.01\,M_\odot\,\mathrm{s}^{-1}$ to $0.07\,M_\odot\,\mathrm{s}^{-1}$. The regions where these clusters form remain gas-rich ($>20\%$ gas mass fraction) up to the point where $90\%$ of the cluster mass has already formed, with the formation process being fuelled by inflowing gas. The gas which is not consumed by star formation is evacuated by supernova explosions.

The cluster stars in the three most massive clusters inherit $\sim 30$\%--$40\%$ of the angular momentum in the gas at the start of the star formation process, and the direction of the cluster rotation is set locally by the conditions in the 
star-forming gas rather than by global rotation. As a result, two of the three most massive clusters end up rotating in the reverse direction with respect to the global angular momentum. In the nine most massive clusters, five end up rotating in the opposite direction (but not directly antiparallel) compared to the global angular momentum. We conclude that the rotational properties of the clusters are directly related to the local angular momentum in the progenitor gas, rather than for example the global rotation or the galactic merger orbit. 

The intrinsic velocity anisotropy within $r_\mathrm{h}$ in the young clusters is slightly radially biased and evolves towards isotropy. The radial distribution of velocity anisotropy in the three most massive clusters is therefore already after 100 Myr of evolution fairly isotropic and in good agreement with observed radial isotropy curves measured in Milky Way globular clusters.

We compare the projected 2D velocity distributions produced for the nine most massive clusters to observations in recent globular cluster surveys. The clusters immediately after formation show higher values of $V_\mathrm{peak}/\sigma_0$ on average compared to the population of old globular clusters, which is consistent with the idea that the rotational rate of star clusters slows down as they evolve even though the majority of star clusters form with non-zero angular momenta. The massive clusters have angular momentum parameters $\lambda_R \lesssim 0.5$ peaking in their edge-on projections. Fits of the line-of-sight velocity distributions  with Gauss-Hermite functions reveal that the third-order coefficients, $h_3$, are anti-correlated with the line-of-sight velocity. This indicates asymmetric line-of-sight velocity distributions with steep leading wings, which have been interpreted as signatures of the dissipative formation process in galaxy formation simulations \citep{2014MNRAS.444.3357N}. The radial profiles of line-of-sight velocity and velocity dispersion measured in the plane of rotation show values comparable to present-day young massive stars clusters as well as observations of old globular clusters.

In hydrodynamical simulation which resolve the internal structure of star clusters, rotation seems to inevitably result due to the conservation of angular momentum in the star-forming gas. As star clusters evolve, they lose mass along with angular momentum. As observational surveys detect rotation in a significant fraction of very old globular clusters, we are beginning to envisage what the rotational properties of these evolutionary remnants might have originally been using simulations without having to resolve their actual formation environment at high redshifts. On the other hand, we currently only have resolved data for massive star clusters in our immediate local neighbourhood. Observations of the resolved velocity structure of less massive star clusters, and especially young star clusters would directly test the simulated results presented in this paper, as
our simulations extend to masses several orders of magnitude below the currently observationally accessible mass range.

\small
\begin{acknowledgements} 
The authors thank Nate Bastian, Sebastian Kamann, Angela Adamo, Rainer Spurzem and Jens Thomas for helpful discussions. N.L. acknowledges the financial support by the Jenny and Antti Wihuri Foundation. TN acknowledges support from the Deutsche Forschungsgemeinschaft (DFG, German Research Foundation) under Germany's Excellence Strategy - EXC-2094 - 390783311 from the DFG Cluster of Excellence "ORIGINS". N.L. and P.H.J. acknowledge support by the  European Research Council via ERC Consolidator Grant KETJU (no. 818930). S.W. acknowledges support by the European Research Council via ERC Starting Grant RADFEEDBACK (no. 679852) and by the German Science Foundation via CRC956, Project C5. The computations were carried out at CSC -- IT Center for Science Ltd. in Finland and at Max-Planck Institute for Astrophysics in Germany.

\end{acknowledgements}

\end{document}